\documentclass[aps,preprint,superscriptaddress]{revtex4}

\usepackage{natbib}
\usepackage[dvips]{graphicx}
\usepackage{amsmath}
\usepackage{amssymb}
\usepackage{multirow}

\newcommand{\grl}{\emph{Geophys. Res. Lett.}}
\newcommand{\jgr}{\emph{Jour. Geophys. Res.}}

\DeclareMathOperator{\sech}{sech}

\newcommand{\yavg}[1]{\overline{#1}}
\newcommand{\tavg}[1]{\widetilde{#1}}
\newcommand{\flavg}[1]{\left<{#1}\right>}

\begin{document}

\title{Drift Turbulence, Particle Transport, and Anomalous Dissipation at the Reconnecting Magnetopause }

\author{A. Le}
\affiliation{Los Alamos National Laboratory, Los Alamos, New Mexico 87545, USA}

\author{W. Daughton}
\affiliation{Los Alamos National Laboratory, Los Alamos, New Mexico 87545, USA}

\author{O. Ohia}
\affiliation{Los Alamos National Laboratory, Los Alamos, New Mexico 87545, USA}

\author{L.-J. Chen}
\affiliation{University of Maryland, College Park, College Park, Maryland 20742 USA and NASA Goddard Space Flight Center, Greenbelt, Maryland 20771 USA}

\author{Y.-H. Liu}
\affiliation{Dartmouth College, Hanover, New Hampshire 03755, USA}

\author{S. Wang}
\affiliation{University of Maryland, College Park, College Park, Maryland 20742 USA and NASA Goddard Space Flight Center, Greenbelt, Maryland 20771 USA}

\author{W. D. Nystrom}
\affiliation{Los Alamos National Laboratory, Los Alamos, New Mexico 87545, USA}

\author{R. Bird}
\affiliation{Los Alamos National Laboratory, Los Alamos, New Mexico 87545, USA}
\begin{abstract}

Using fully kinetic 3D simulations, the reconnection dynamics of asymmetric current sheets are examined at the Earth's magnetopause. The plasma parameters are selected to model MMS magnetopause diffusion region crossings with guide fields of 0.1, 0.4, and 1 of the reconnecting magnetosheath field. In each case, strong drift-wave fluctuations are observed in the lower-hybrid frequency range at the steep density gradient across the magnetospheric separatrix. These fluctuations give rise to cross-field electron particle transport.  In addition, this turbulent mixing leads to significantly enhanced electron parallel heating in comparison to 2D simulations. We study three different methods of quantifying the anomalous dissipation produced by the drift fluctuations, based on spatial averaging, temporal averaging, and temporal averaging followed by integrating along magnetic field lines. Comparison of the different methods reveals complications in identifying and measuring the anomalous dissipation. Nevertheless, the anomalous dissipation from short wavelength drift fluctuations appears weak for each case, and the reconnection rates observed in 3D are nearly the same as in 2D models. The 3D simulations feature a number of interesting new features that are consistent with recent MMS observations, including cold beams of magnetosheath electrons that penetrate into the hotter magnetospheric inflow, the related observation of decreasing temperature in regions of increasing total density, and an effective turbulent diffusion coefficient that agrees with predictions from quasi-linear theory.
\end{abstract}

\maketitle

\section{Introduction}

Magnetic reconnection plays a central role in the dynamics of the magnetosphere, the solar wind, the solar corona, and a variety of other space plasmas where particle collisions are negligible \cite{priest:2002,yamada:2010}. In these collisionless plasmas, kinetic effects ultimately control the evolution of reconnecting current sheets. Observing the small-scale layers, typically referred to as diffusion regions \cite{mozer:2009grl,graham:2014,hesse:2014,hesse:2016}, where kinetic effects are essential is a main goal of NASA's MMS mission \cite{moore:2013}. In the first phase of its mission, MMS collected high-resolution observations of electron diffusion regions near Earth's magnetopause, where reconnection allows solar wind plasma to couple to and enter the Earth's magnetosphere \cite{sonnerup:1974,paschmann:1979,russell:1979,cowley:1982,wygant:1983,phan:2000,mozer:2002}. Reconnection at the magnetopause is characterized by strong asymmetries in the plasma and magnetic field conditions on either side \cite{cassak:2007,borovsky:2007,mozer:2008,pritchett:2008,pritchett:2009}, with the magnetosheath side containing high-density shocked solar wind plasma and the opposite magnetospheric side having a stronger magnetic field and hotter, more tenuous plasma. Depending on the orientation of the solar wind magnetic field, magnetopause reconnection may also occur with a variety of guide magnetic fields that reduce the total magnetic shear across the magnetopause current layer. 

Fully kinetic simulations have been instrumental in understanding the details of the reconnection process in the complicated asymmetric geometry typical of the magnetopause, and the simulations have guided the interpretation of the data collected by spacecraft. So far, the vast majority of the simulations have been performed in 2D geometries, which have nonetheless captured many of the key field and particle signatures that have been observed in MMS data of magnetopause reconnection \cite{bessho:2016,chen:2016,egedal:2016,shay:2016,torbert:2016,burch:2016science,burch:2016grl,zenitani:2017,egedal:2018}.

Two-dimensional simulations, however, preclude the development of an important class of instabilities. Of particular importance at the magnetopause are drift instabilities driven by the diamagnetic current carried by either electrons or ions at the steep density gradient between the magnetospheric and magnetosheath plasmas. The fastest growing modes of the lower-hybrid drift instability (LHDI) are approximately electrostatic \cite{davidson:1975}, with typical wavelengths of order $k\rho_e\sim1$ and frequencies of $\omega\sim\omega_{LH} = \omega_{pi}/\sqrt{(1+\omega_{pe}^2/\omega_{ce}^2)}$ \cite{daughton:2003}. The mode is expected to be unstable under typical magnetospheric plasma conditions and has indeed been observed near magnetopause reconnection sites \cite{vaivads:2004, graham:2017} and in laboratory reconnection experiments \cite{carter:2001,yoo:2014,yoo:2017}. 

Researchers have long speculated that LHDI fluctuations may contribute to anomalous transport at the magnetopause. Lower-hybrid range fluctuations have been found to have electric potentials of amplitude $e\delta\phi/T_e\sim0.1$ at both the magnetopause \cite{bale:2002} and in the geomagnetic tail \cite{norgren:2012}, suggesting that they could strongly interact with the electrons. This wave-particle interaction can lead to anomalous cross-field particle transport \cite{davidson:1975} that carries magnetosheath plasma into the magnetosphere even in the absence of magnetic reconnection \cite{treumann:1991,gary:1990}, and this particle mixing was observed in 3D kinetic simulation with weak guide field\cite{le:2017}. 

Important for reconnection, which requires a breaking of the electron frozen-in condition, the lower-hybrid fluctuations may also couple electrons and ions to produce an anomalous resistivity \cite{huba:1977}. Because LHDI is weakened at higher plasma $\beta$ (ratio of thermal pressure to magnetic pressure) \cite{davidson:1977} and takes on a longer wavelength electromagnetic character in the center of thin current sheets \cite{daughton:2003}, it was not expected to contribute anomalous resistivity at the X-line during reconnection (where the in-plane magnetic field has a null). Indeed, in 3D simulations of reconnection with density asymmetries \cite{roytershteyn:2012}, the electrostatic LHDI was found to be localized to the density gradient region away from the X-line. More recent simulations \cite{le:2017,price:2016,price:2017}, however, with very strong density and temperature asymmetries matching an MMS event with weak guide field \cite{burch:2016science}, found that the electrostatic drift fluctuations could transiently penetrate into the X-line region. In addition, slower-growing electromagnetic fluctuations may persist and contribute to dissipation through a so-called anomalous viscosity \cite{che:2011,price:2016}.

We re-examine the importance of drift fluctuations for anomalous transport and dissipation near magnetopause reconnection sites with a set of fully kinetic 3D simulations that are based on parameters from three MMS diffusion region encounters with varying guide magnetic fields \cite{burch:2016science,chen:2017,burch:2016grl}. The paper is outlined as follows: In Section \ref{sec:pic}, we present the parameters and set-up of our fully kinetic particle-in-cell calculations. The cross-field particle transport and enhanced parallel electron heating induced by lower-hybrid range fluctuations are analyzed in Section \ref{sec:parttransport}. We consider three different methods of quantifying the anomalous dissipation in the 3D simulations in Section \ref{sec:ohm}, and a summary discussion follows.

\section{PIC Simulations}
\label{sec:pic}

We performed three 3D PIC simulations of reconnecting asymmetric current sheets using the fully kinetic particle-in-cell code VPIC \cite{bowers:2008}. In order to use parameters relevant to Earth's magnetopause, the upstream plasma conditions on each side of the asymmetric current sheet were selected to mimic magnetopause conditions from three MMS diffusion region encounters with guide magnetic fields of $\sim0.1$ \cite{burch:2016science} (some results from this run appear in \cite{le:2017}), $\sim0.4$ \cite{chen:2017}, and $\sim1$ \cite{burch:2016grl} of the magnetosheath reconnecting magnetic field component. While velocity shear across the magnetopause is common \cite{nakamura:2017}, we do not include bulk flows in our initial conditions. For feasibility, each simulation employs a reduced ion-to-electron mass ratio of $m_i/m_e=100$ and a reduced electron plasma-to-cyclotron ratio $\omega_{pe0}/\omega_{ce0}$ (see Table 1). Each computational domain is $L_x\times L_y\times L_z = 4096\times1024\times2048$ cells $=40d_i\times10d_i\times20d_i$ (with $d_i$ based on the initial magnetosheath density). The boundary conditions are periodic in $x$ and $y$, and conducting for fields and reflecting for particles in $z$. Macroparticles from the high-density magnetosheath side and the low-density magnetosphere side are loaded as separate populations with different numerical macroparticle weights. This also allows us to quantify the mixing of particles originating from opposite sides of the magnetopause. In addition, the weights are selected so that the low-density magnetosphere plasma and the higher-density magnetosheath plasma are each resolved with $\sim150$ particles per species per cell, yielding $\sim2.6\times10^{12}$ total numerical particles. For asymmetric layers with a larger density jump (up to a factor of $\sim16$ in this paper), this approach is crucial for limiting the numerical noise.


The initial conditions include a Harris sheet current-carrying population superposed on an asymmetric Maxwellian background \cite{roytershteyn:2012}. In particular, the initial magnetic field has components
\begin{eqnarray}
B_x &=& \frac{1}{2}\times\left[(B_1-B_0) + (B_1+B_0)\tanh(\frac{z}{\lambda})\right]\\ \nonumber
B_y &=& B_g\\ \nonumber
B_z &=& 0
\end{eqnarray}
where we use subscript 0 to refer to upstream magnetosheath quantities and subscript 1 to refer to upstream magnetosphere quantities and the initial current sheet width is chosen as $\lambda=1d_i$ ($d_i$ is the ion inertial length based on the initial magnetosheath density). Each background species (electrons and ions) is loaded as a Maxwellian with the following asymmetric temperature profile:
\begin{equation}
T_s = \frac{1}{2}\times\left[(T_{s1}+T_{s0}) + (T_{s1}-T_{s0})\tanh(\frac{z}{\lambda})\right]\\ \nonumber
\label{eq:T}
\end{equation}
The total plasma density profile is then selected to establish gross single-fluid hydrodynamic pressure balance: 
\begin{equation}
n =  \frac{(p_{1}+p_{0}) + (p_{1}-p_{0})\tanh(\frac{z}{\lambda})}{(T_{1}+T_{0}) + (T_{1}-T_{0})\tanh(\frac{z}{\lambda})}+ n_H\sech^2(z/\lambda)
\end{equation}
where the total temperature $T_k = T_{ek}+T_{ik}$ and total pressure $p_k = n_k(T_{ek}+T_{ik})$ for $k=0,1$; the Harris density is $n_H = \mu_0(B_0+B_1)^2/8T_H$; and we choose the Harris population temperature of each species $s=i,e$ to equal the hotter magnetospheric temperature: $T_{sH}=T_{s1}$. While the initial conditions are not an exact Vlasov equilibrium, the system rapidly settles into a quasi-equilibrium within a few ion gyroperiods \cite{pritchett:2008,roytershteyn:2012}, which includes the development of a strong electrostatic field normal to the current sheet that confines the magnetosheath ions and generates a strong electron $E\times B$ flow \cite{pritchett:2008,price:2017}. Reconnection is seeded with a magnetic perturbation that generates a single X-line at the center of the current sheet extended in the $y$-direction. The reconnection rate typically peaks by a time of $t\sim20/\omega_{ci}$. Unless otherwise noted, we plot data from times in the interval $t\sim(30$---$35)/\omega_{ci}$ during a period of quasi-steady reconnection. In Figs.~\ref{fig:runs}(a-c), we plot the fully evolved current density from each 3D PIC simulation. 

The fluctuations along the boundary layer separating the magnetospheric plasma from the reconnection exhaust have been identified as LHDI in previous work. They have a frequency near the lower-hybrid frequency and wave vectors with ${\bf{k\cdot B}}\sim0$ and $k\rho_e\sim1$ \cite{price:2016,price:2017,le:2017}. In Run A with a weak guide field, there is also a large-scale kink of the current sheet with the lowest mode number that fits in the simulation domain. This is likely the electromagnetic \cite{daughton:2003} branch of LHDI with $k\sqrt{\rho_e\rho_i}\sim1$. It is stabilized by a guide magnetic field, and any kinking of the current sheet in Runs B and C is substantially weaker.

\begin{table}
\caption{Physics parameters used for VPIC simulations. The density and temperature ratios along with the plasma $\beta$ values match conditions of the diffusion regions encountered by MMS in the references of the last column. For computational feasibility, an artificially reduced value of $\omega_{pe0}/\omega_{ce0}$ and a reduced ion-to-electron mass ratio of $m_i/m_e$ are employed.}
\begin{center}
  \begin{tabular*}{15cm}{@{\extracolsep{\fill} }  l *{9}{|c}}
    \hline\hline
    Run & $B_g/B_0$ & $\beta_0$ & $n_1/n_0$ & $B_1/B_0$ & $T_{i0}/T_{e0}$ & $T_{i1}/T_{i0}$ & $T_{e1}/T_{e0}$ & $\omega_{pe0}/\omega_{ce0}$ & Ref.\\ \hline\hline
    A & 0.099 & 3.0 & 0.062 & 1.7 & 11  & 6.3 & 3.3 & 1.5 & \cite{burch:2016science}\\ \hline
    B & 0.41  & 1.3 & 0.15  & 1.3 & 11  & 3.2 & 3.8 & 1.5 & \cite{chen:2017}\\ \hline
    C & 1.0   & 6.2 & 0.18  & 2.0 & 8.3 & 3.0 & 1.7 & 2.0 & \cite{burch:2016grl}\\ \hline
    \hline
  \end{tabular*}
\end{center}
\end{table}

\section{Particle Transport and Heating}
\label{sec:parttransport}

\begin{figure}
\includegraphics[width = 18.0cm]{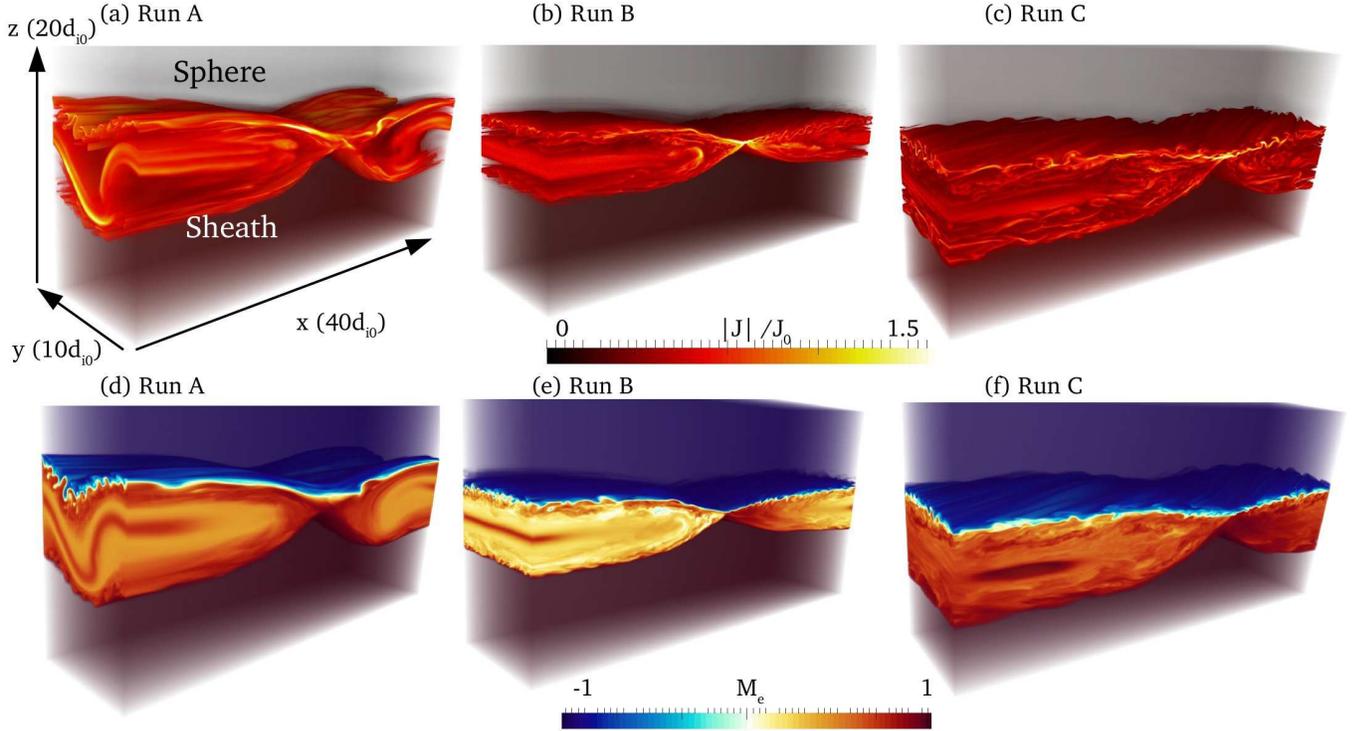}
\caption{Volume rendering of current density and electron mixing $M_e$ from each 3D simulation. Each run here is shown at $t=30/\omega_{ci}$.\label{fig:runs}}
\end{figure}

As noted in the Introduction, the anomalous transport produced by LHDI across steep density gradients was considered as a possible mechanism for regulating the thickness of Earth's magnetopause boundary layer \cite{treumann:1991,gary:1990}. Even in the absence of magnetic reconnection, drift fluctuations may transport plasma from the dense magnetosheath across the magnetic field into the relatively tenuous magnetosphere. Here, we study the particle mixing of magnetosheath and magnetosphere plasma in our 3D kinetic simulations. The particles on opposite sides of the initial magnetopause current sheet are tagged separately, allowing their mixing over time to be diagnosed. We use the mixing measure introduced in Ref.~\cite{daughton:2014} defined as:
\begin{equation}
M_e = \frac{n_{sh}-n_{sp}}{n_{sh}+n_{sp}}
\label{eq:me}
\end{equation}
where $n_{sh}$ is the local number density of electrons that orginated in the magnetosheath and $n_{sp}$ is the density of magnetosphere electrons. The mixing measure $M_e$ varies from $1$ where all electrons originate in the magnetosheath to $-1$ where all electrons are from the magnetosphere. A 3D volume rendering of the mixing measure $M_e$ from each 3D simulation is plotted in Figs.~\ref{fig:runs}(d-f). The electron populations are strongly mixed within the region of LHDI fluctuations along the magnetospheric separatrix.

\begin{figure}
\includegraphics[width = 16.0cm]{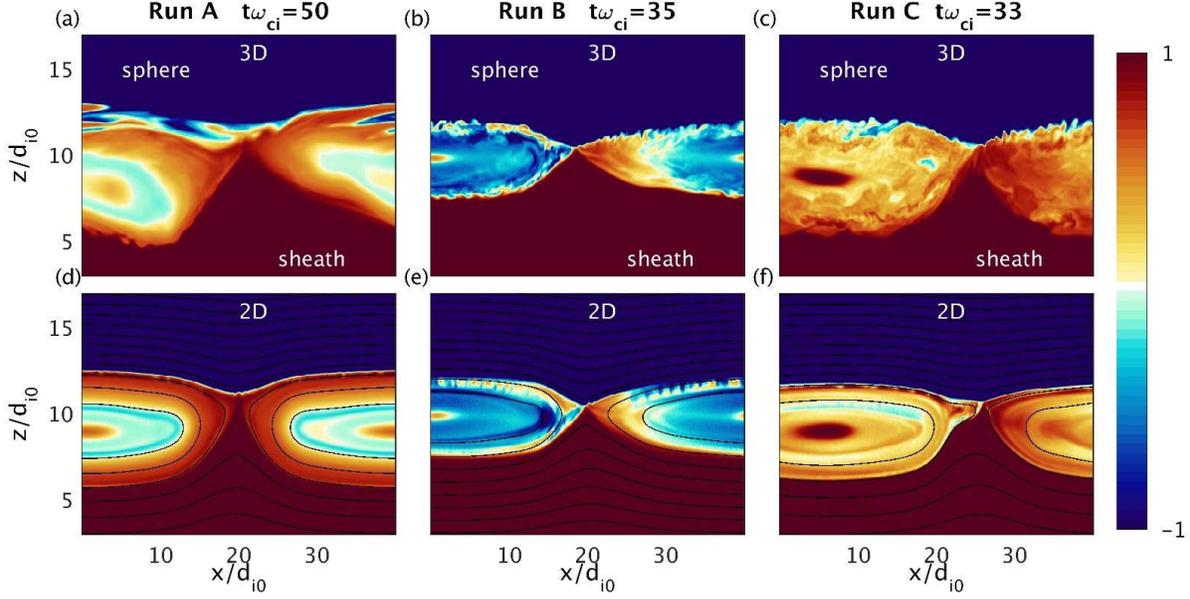}
\caption{$M_e$. Comparison of electron mixing diagnostic $M_e$ introduced in Ref.~\cite{daughton:2014} in each 2D simulation and from a slice of each 3D simulation. Data here are from the last available time step of each run. \label{fig:emix}}
\end{figure}

\begin{figure}
\includegraphics[width = 12.0cm]{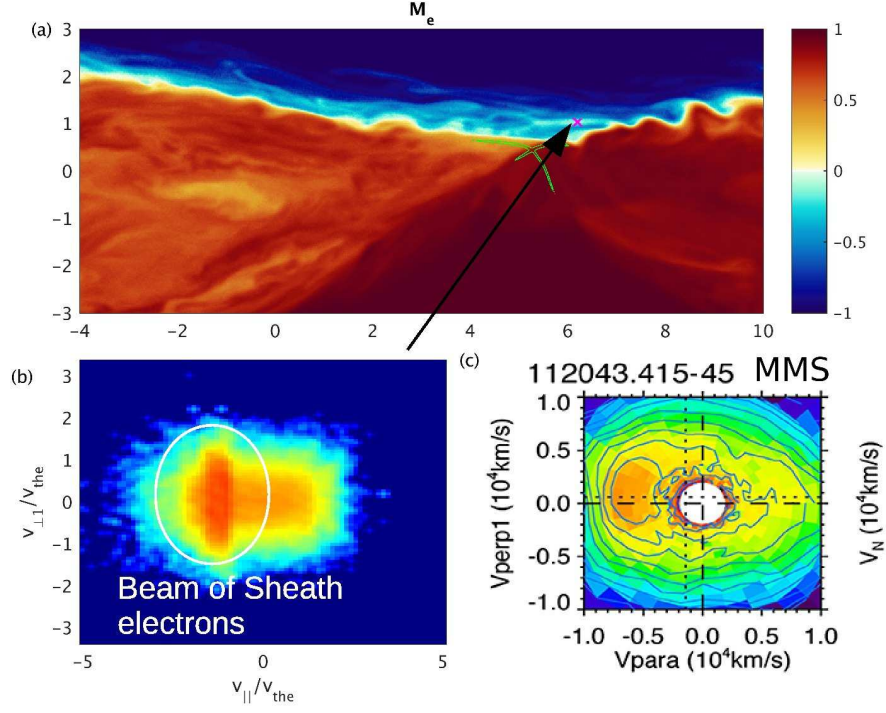}
\caption{(a) Mix measure $M_e$ near the approximate X-line (indicated by green contours) in Run C with $B_g/B_0\sim1$. (b) A reduced electron distribution in $v_\parallel-v_\perp$ space at the point marked $\times$ in panel (a), which exhibits a parallel beam of magnetosheath electrons. (c) A slice from a similar electron distribution from MMS observations \cite{burch:2016grl} of a magnetopause reconnection event with parameters modeled by this simulation. For each distribution, $v_{\perp1}$ is in the direction of the mean perpendicular electron flow. \label{fig:beam}}
\end{figure}

In Fig.~\ref{fig:emix}, $M_e$ is plotted from each 3D simulation in the top panels and from a corresponding 2D simulation in the lower panels. For the 2D simulations in Figs.~\ref{fig:emix}(d-f), sample in-plane magnetic field lines, which are contours of the out-of-plane magnetic vector potential $A_y$ are also indicated. Note that because the canonical momentum $p_y = mv_y-eA_y$ is strictly conserved in 2D, each electron never moves farther than an in-plane Larmor radius from a single magnetic flux surface. The particle mixing is therefore limited to an electron orbit width across the magnetic separatrices \cite{egedal:2016}, resulting in the sharp boundaries in $M_e$ across the magnetopause surface. In a system with 3D fluctuations, however, no conservation property prohibits electrons from being transported across the magnetic separatrices \cite{le:2017}. In fact, in Figs.~\ref{fig:emix}(a-c), there is a region of strong magnetosheath and magnetosphere mixing particularly along the magnetospheric separatrix where the density gradient and LHDI fluctuations are strongest. There is evidence of the electron mixing in MMS observations of magnetopause reconnection sites. In particular, a population of lower-energy electrons presumed to be sourced from the magnetosheath have been observed in the magnetosphere inflow region in conjunction with LHDI fluctuations at the current density layer \cite{graham:2017}. In addition, beams of electrons, presumed to be sourced from the magnetosheath, have been observed in the magnetospheric inflow. An example of a parallel beam of magnetosheath electrons (identified by their particle tag in the simulation) is plotted in Fig.~\ref{fig:beam}(b) from the point labeled $\times$ in Fig.~\ref{fig:beam}(a). In this particular distribution, the magnetosheath electron beam is moving towards the X-line ($v_\parallel<0$), although distributions from nearby regions display magnetosheath beams moving away from the X-line. A similar distribution measured by MMS [See Fig. 8 of Ref. \cite{burch:2016grl}] is plotted in Fig.~\ref{fig:beam}(c) from the event with density and temperature asymmetry matching this simulation, where the parallel beam of high phase-space density is presumed to be composed of electrons that originated in the magnetosheath.

\subsection{Quasi-Linear Estimates}

The enhanced particle mixing observed in 3D is a direct result of out-of-plane fluctuations. As a first method to quantify the anomalous or fluctuation-induced particle transport across the magnetospheric separatrix, we employ spatial averaging in the $y$ direction \cite{che:2011,roytershteyn:2012,price:2016,che:2017,le:2017}. We split each quantity $Q$ into mean and fluctuating components defined as $Q = \yavg{Q}+ \delta Q$, with the mean component given by the $y$-average over the simulation domain:
\begin{equation}
\yavg{Q}(x,z,t) = \frac{1}{L_y}\int_0^{L_y} Q(x,y,z,t) dy,
\end{equation}
and where the fluctuating component satisfies $\yavg{\delta Q} = 0$ by construction. Note that for pairs of quantities $Q$ and $R$, $\yavg{QR} = (\yavg{Q})(\yavg{R}) + \yavg{\delta Q\delta R}$, and the anomalous particle transport is ascribed to the correlated fluctuations in density and bulk flow, $\yavg{\delta n\delta{\bf{u}}}$.

\begin{figure}
\includegraphics[width = 12.0cm]{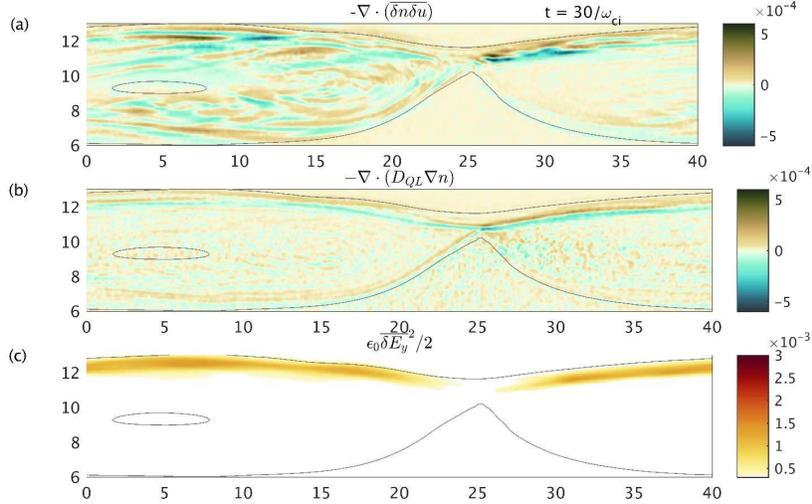}
\caption{(a) Fluctuating particle transport across the magnetospheric separatrix in Run C ($B_g/B_0\sim1$). (b) Quasi-linear estimate for the particle transport induced by LHDI. (c) The fluctuation amplitude of the electric field directly evaluated from the simulation data. Gray lines mark where the electron mix measure $M_e=\pm0.9$.\label{fig:qldiff}}
\end{figure}

In the quasi-linear approximation, the fluctuating density $\delta n$ and  electron drift $\delta{\bf{u}}$ are computed for LHDI from linear instability theory. The density profile is then predicted to relax through a diffusion equation of the form $\partial \yavg{n}/\partial t\sim \nabla\cdot(D_{QL}\nabla\yavg{n})$, where the anomalous particle flux $\yavg{\delta n\delta{\bf{u}}}$ is proportional to the gradient of the mean density profile $\nabla\yavg{n}$ through a diffusion coefficient $D_{QL}$.  For LHDI, the quasi-linear diffusion coefficient $D_{QL}$ is given by \cite{treumann:1991,winske:1995}
\begin{equation}
D_{QL} \sim 0.5\rho_e^2\nu_{QL}(1+\frac{T_i}{T_e})
\end{equation}
where the effective anomalous collision frequency is
\begin{equation}
\nu_{QL} \sim 0.6 \omega_{lh}\left(1+\frac{\omega_{pe}^2}{\omega_{ce}^2}\right)\frac{m_i}{m_e}\frac{E_f}{nT_i}
\label{eq:dql}
\end{equation}
and $E_f ~\sim \epsilon_0\yavg{(\delta E_y)^2}/2$ is the energy density of the fluctuating electric field. The energy density of the fluctuating electric field $E_f$ in the saturated state is estimated to be \cite{winske:1978}
\begin{equation}
E_f\sim \frac{nm_eV_d^2}{4(1+\omega_{pe}^2/\omega_{ce}^2)}
\label{eq:esat}
\end{equation}
given in terms of the relative drift velocity $V_d$ between the electrons and the ions. This estimate agrees with the peak fluctuation amplitudes in the simulations within a factor of order unity. The spatial dependence of the fluctuation amplitude, however, depends also on the details of the drift velocity, magnetic field, and density profiles.  In order to compare the quasi-linear estimate to the simulations, we evaluate the average fluctuation amplitude $\yavg{\delta E}^2$ directly from the simulation [see Fig.~\ref{fig:qldiff}(c)]. An example from Run C with $B_g/B_0\sim1$ is plotted in Figures \ref{fig:qldiff}(a-b), and we find reasonable agreement between the fluctuating particle flux and the quasi-linear theoretical estimate.

The coefficient $D_{QL}$ has been estimated directly from observations by time-filtering multi-spacecraft measurements of the ratio $\yavg{\delta n\delta{\bf{u}}}/\nabla\yavg{n}$ \cite{vaivads:2004}, and it was found to agree with theoretical predictions. In MMS observations of a magnetopause reconnection event \cite{graham:2017}, a peak value of $D_{QL}\sim0.8\times10^9$ m/s$^2$ was found. Based on a typical fluctuating electric field of $|\delta E|\sim 20$ mV/m for the same event, we find $D_{QL}\sim0.6\times10^9$ m/s$^2$ following from Eq.~\ref{eq:dql}. Taking an average drift velocity $V_d\sim300$ km/s yields $D_{QL}\sim0.1\times10^9$ m/s$^2$ using Eq.~\ref{eq:esat}, which is also in rough agreement with the observed values, though smaller than the peak value.

\begin{figure}
\includegraphics[width = 12.0cm]{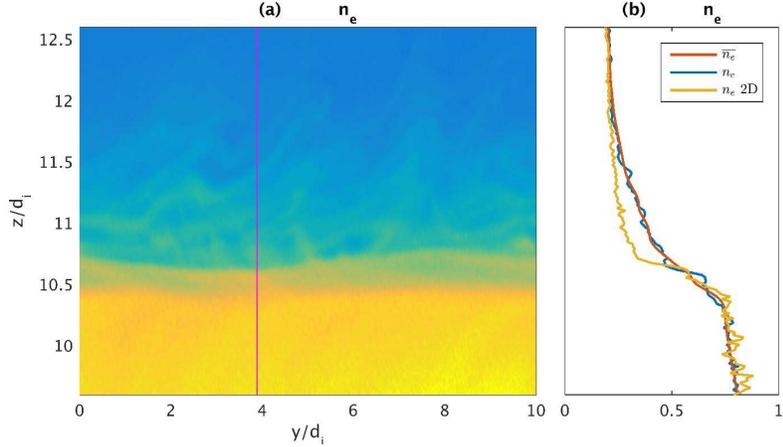}
\caption{(a) Cut at the X-line of the electron density in the $y-z$ plane from Run C ($B_g/B_0\sim1)$ at $t\omega_{ci}=30$. (b) Line out of the density profile across the current sheet along the cut indicated in (a) along with corresponding cuts of the $y$-averaged density profile, and the density profile from a 2D simulation. Particle transport across the magnetospheric separatrix relaxes the density gradient.\label{fig:ne2d3d}}
\end{figure}

Because of the direction of the particle transport shown in Figs.\ref{fig:qldiff}(a-b), the main effect of the turbulent transport is to relax the density gradient across the magnetospheric separatrix. This process thus carries plasma from the high density magnetosheath plasma into the magnetospheric inflow region. Figure~\ref{fig:ne2d3d} compares the 3D density profile across the current sheet to the $y$-averaged and the 2D density profiles. In agreement with Ref.~\cite{price:2017}, the 3D density profile is broadened compared to 2D. 

\subsection{Magnetic Field Line Mixing vs. Cross-field Transport}

\begin{figure}
\includegraphics[width = 16.0cm]{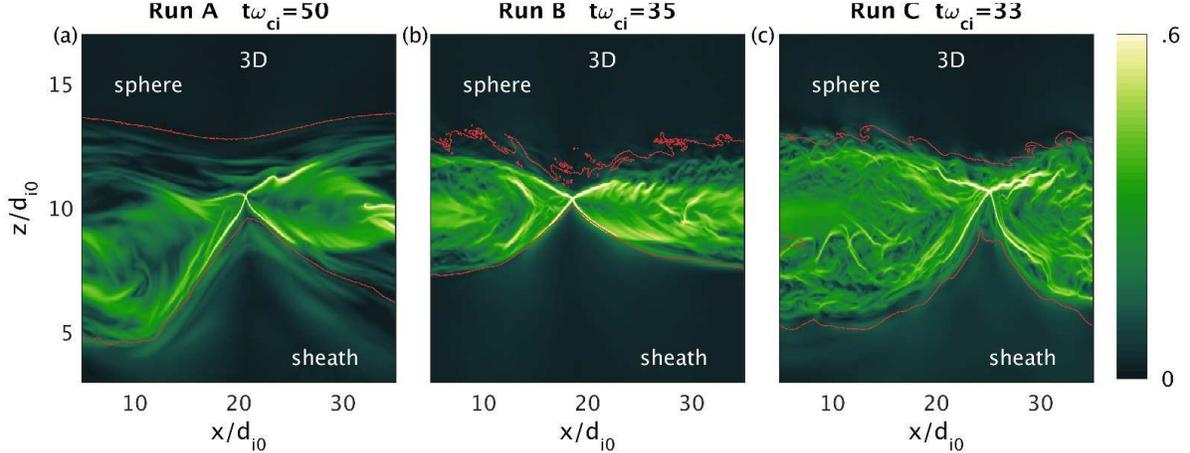}
\caption{The magnetic field exponentiation factor $\sigma$ \cite{daughton:2014} from a plane of seed points in each 3D simulation. Plotted here is $max(\sigma_b,\sigma_f)$, where each $\sigma$ is calculated by tracing either $f$orward or $b$ackward along the magnetic field lines. The ridges of large values highlight the quasi-separtrix layer (QSL), which is an approximate X-line and pair of separatrices.  The red contour in each panel marks where the electron mix measure $M_e=0.99$.  \label{fig:ftle}}
\end{figure}

Within the quasi-linear theory of transport by electrostatic LHDI, the fluctuating electron drift results from ${\bf{E}}\times{\bf{B}}$ motions across the magnetic field. The magnetosheath electrons within the turbulent mix layer along the magnetospheric separatrix are mixed by cross-field transport, rather than by parallel streaming along mixed magnetic field lines. To verify this, we study the magnetic geometry and topology. 

In guide field regime (without magnetic nulls), well-defined topological boundaries between reconnected magnetic field domains may not exist. In this limit, the concept of quasi-separatrix layers (QSLs) was introduced to identify regions where the magnetic field lines very rapidly diverge \cite{priest:1995,titov:2002}, which coincide with topological separatrices when applied to 2D systems. The QSLs may be calculated with the so-called squashing factor \cite{titov:2002}, a geometric measure computed from the map that takes initial seed points to final points a finite distance away along the magnetic field. To locate an approximate QSL in our runs, we compute a closely related measure, the exponentiation factor $\sigma$ \cite{daughton:2014,borgogno:2011}, which is also calculated by integrating along magnetic field lines. We trace magnetic field lines a distance of $L=5d_{i0}$ in the forward (along $\bf{B}$) and backward (along $-\bf{B}$) directions from each point $\bf{x_0}$ within a seed plane to some final point $\bf{x_f}$. The factor $\sigma$ is then defined as $\sigma=\log\sqrt{\lambda_m}$, where $\lambda_m$ is the maximal eigenvalue of the symmetric matrix
\begin{equation}
\left(\frac{\partial {\bf{x_f}}}{\partial {\bf{x_0}}}\right)^T\left(\frac{\partial {\bf{x_f}}}{\partial {\bf{x_0}}}\right).
\end{equation}
The QSL may be visualized as the region of large values of $\sigma$. In Fig.~\ref{fig:ftle}, we plot the maximum of the forward and backward $\sigma$ values. In each case, a ridge of large values is found near the approximate X-line, highlighting the QSLs within the reconnection region branching out from the approximate X-line. 

Also plotted in Fig.~\ref{fig:ftle} are contours where the electron mix measure $|M_e|=0.99$, meaning that beyond these boundaries there is practically no mixing of magnetosheath and magnetosphere electrons. Particularly on the magnetosphere side at weaker guide fields, the electron mixing contour is separated from the QSL, confirming that there is relatively little mixing of magnetic field lines although there is particle mixing. This is as opposed to previous 3D simulations of asymmetric reconnection \cite{daughton:2014}, where the electron mixing was mainly attributed to parallel streaming along reconnected field lines, and the electron mixing contour was a good proxy for the magnetic separatrix, as demonstrated by field line tracing. 

\begin{figure}
\includegraphics[width = 16.0cm]{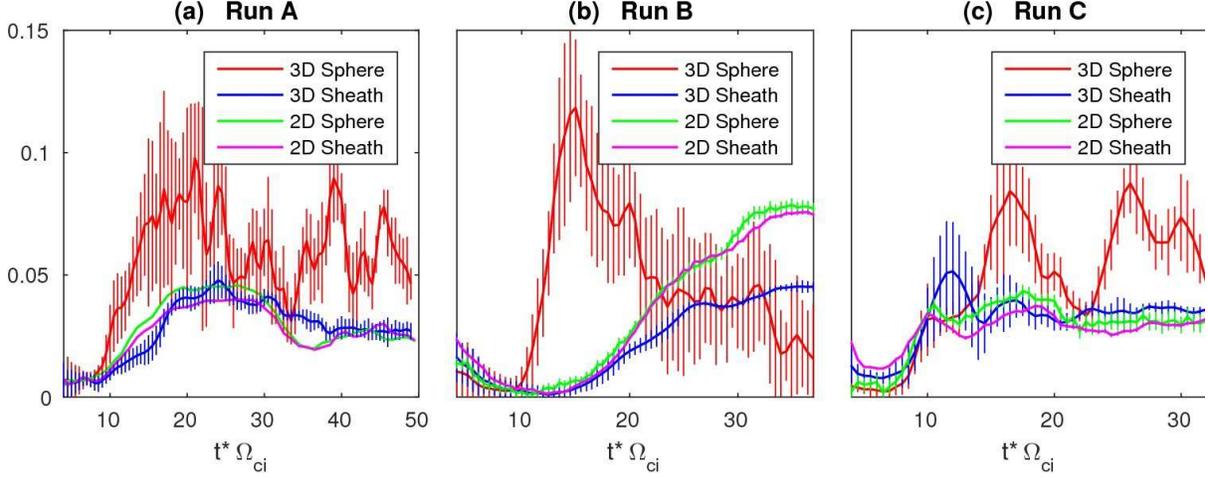}
\caption{Comparison of inferred reconnection rates over the course of each 2D and 3D simulation evaluated with the electron mixing diagnostic introduced in Ref.~\cite{daughton:2014}. The method cannot be used to determine a magnetic reconnection rate on the magnetospheric side because turbulence transports electrons across the magnetospheric separatrix. On the magnetosheath side, however, there is little cross-field particle transport, and the rate is similar in 2D and 3D simulations (except that Run B developed a large secondary island in 3D at $t\sim20/\omega_{ci}$ that reduced the global rate). The 3D fluctuations do not strongly modify the overall reconnection rate. The "error bars" are the variance of the rate measure over a set of cuts at different $x$ values along the length of the current sheet (see Appendix of \cite{daughton:2014} for detailed explanation). \label{fig:rates}}
\end{figure}

In the absence of strong cross-field electron transport, the reconnection rate may be computed by calculating the change in flux of the reconnecting magnetic field in the region of no electron mixing \cite{daughton:2014}. We apply this diagnostic to our runs, and the results are plotted in Fig.~\ref{fig:rates}, where the rates are normalized using a hybrid magnetic field and Alfven speed based on values on both sides of the asymmetric layer following the conventions of Ref.~\cite{cassak:2007}. As expected, the diagnostic is not suitable near the magnetosphere separatrix, where fluctuation-induced particle transport carries magnetosheath electrons across the magnetic separatrix. On the magnetosheath side, however, cross-field electron transport is weak, and we expect the diagnostic to give a good estimate of the global reconnection rate. In fact, the global reconnection rates measured in this way are nearly identical in 2D and on the magnetosheath side in 3D, with normalized rates of $\sim 0.05$ \cite{liu:2017}. This suggests that while the strong turbulence that develops along the magnetosphere separatrix affects plasma transport across the magnetopause, it ultimately has little effect on the overall rate of reconnection. 

\subsection{Enhanced Parallel Heating}

\begin{figure}
\includegraphics[width = 16.0cm]{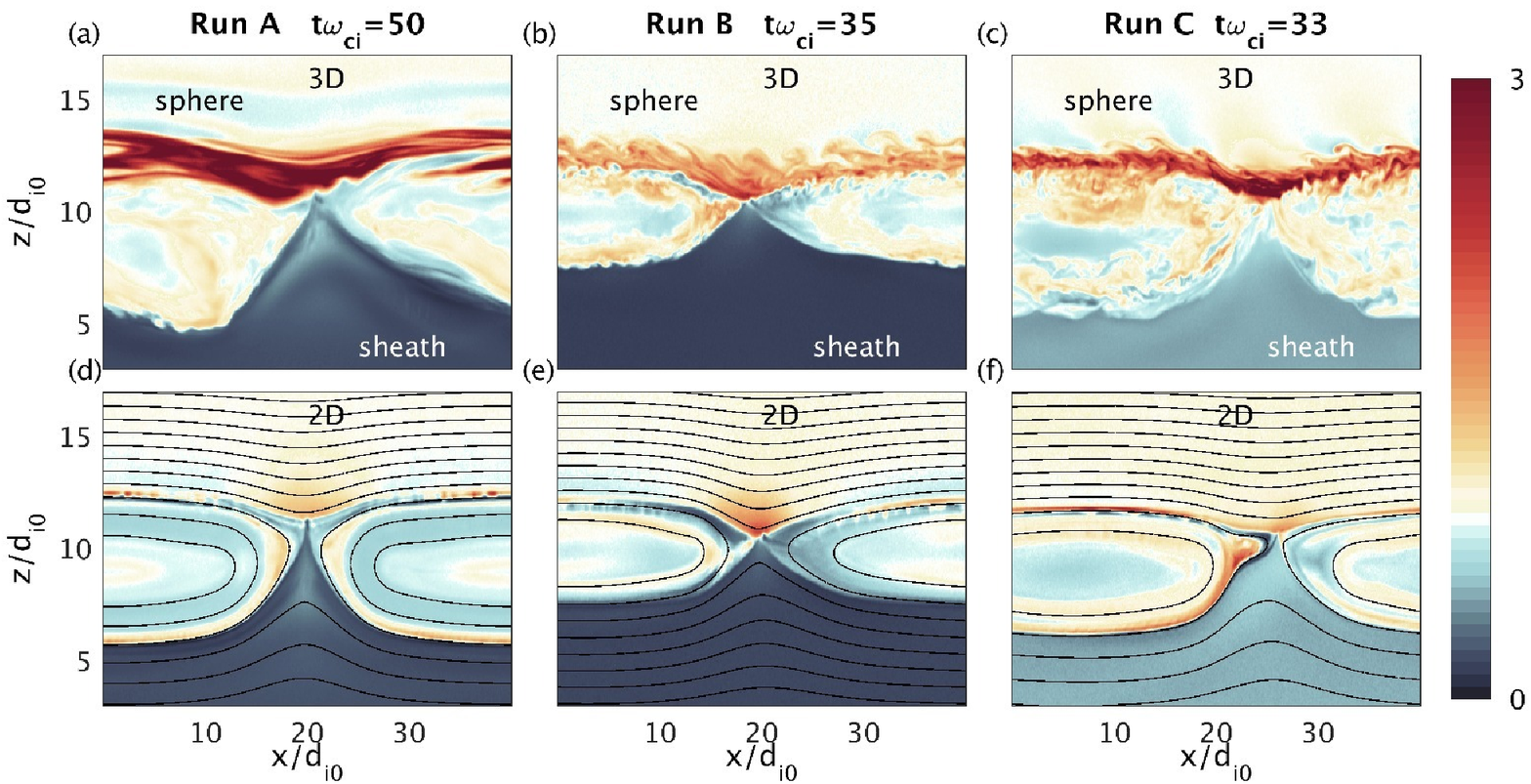}
\caption{Comparison of parallel electron temperature in each 2D simulation and from a slice of each 3D simulation. Note the enhanced $T_{e\parallel}$ within the mix layer near the magnetospheric separatrix in each 3D simulation. \label{fig:tepar}}
\end{figure}

For the weak guide field case, it was previously demonstrated that the particle mixing region also exhibits enhanced heating of the electrons in the direction parallel to the magnetic field \cite{le:2017}. In Fig.~\ref{fig:tepar}, the parallel electron temperature $T_{e\parallel}$ is plotted from each of the three 3D simulations. In all three cases, similar enhancement of $T_{e\parallel}$ is observed within the fluctuation mix layer. In Ref.~\cite{le:2017}, the locally sourced electrons in the inflow regions were found to obey relatively simply equations of state \cite{le:2009, egedal:2011pop,egedal:2013pop} that resemble the CGL \cite{chew:1956} scalings $T_{e\parallel}\propto n^2/B^2$ and $T_{e\perp}\propto B$. The increase of temperature with density implied that the heating was essentially caused by an adiabatic compression of the electron fluid. The simple scaling, however, only applied to electron populations in each inflow region that had remained  on a given side of the separatrix, not those that mixed from the opposite side.

\begin{figure}
\includegraphics[width = 12.0cm]{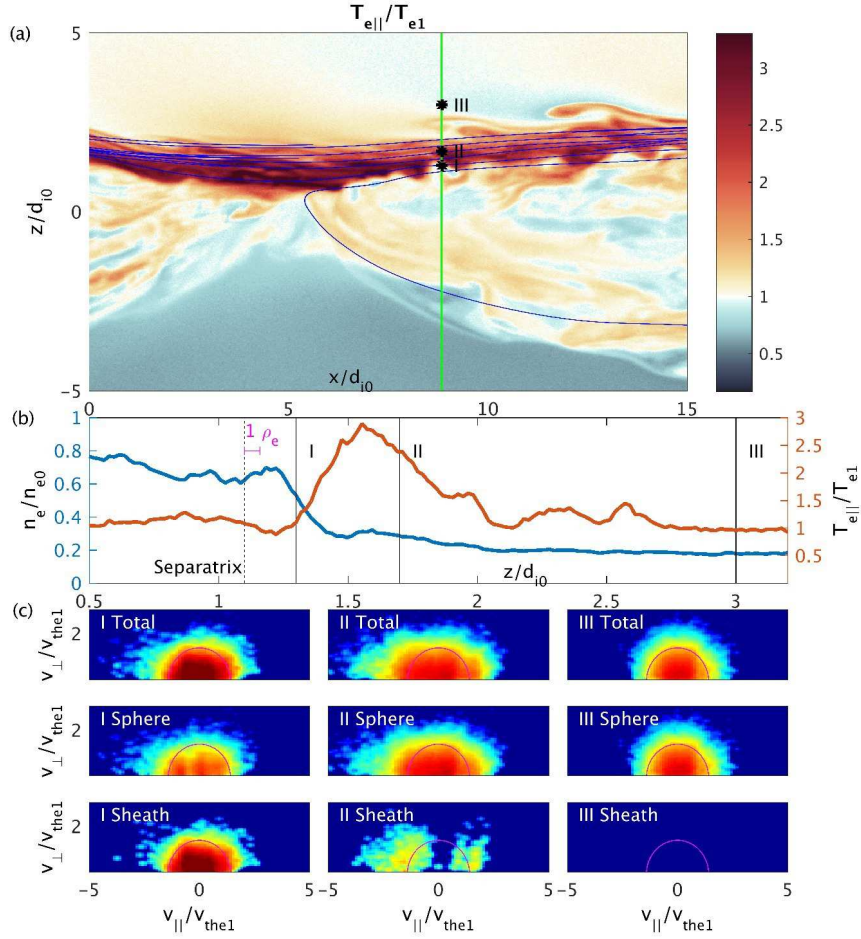}
\caption{(a) Parallel electron temperature from Run C with order unity guide field. The blue contours are in-plane projections of magnetic field lines. (b) Line outs of the density and parallel electron temperature along the $z$ cut marked in green in (a). The location of the last un-reconnected magnetospheric field line is marked by the dashed line. (c) Reduced electron velocity distribution functions in $v_\parallel$-$v_\perp$ space from the points labeled $I$, $II$, and $III$ in (a-b). The top row shows the total electron distribution, the middle row shows only particles that originated on the magnetosphere side, and the bottom row includes only electrons from the magnetosheath. Circular magenta contours are for reference to aid in seeing anisotropy. \label{fig:mixte}}
\end{figure}

A survey of MMS observations of magnetopause reconnection sites found that in most cases a region exists in the magnetospheric inflow where the electron parallel temperature falls even while the density increases \cite{wang:2017}. Because this region was found to cover length scales larger than the electron Larmor radius, the discrepancy with the fluid equations of state predictions was attributed to anomalous particle mixing. We confirm that these signatures are present in our 3D simulations. Figure~\ref{fig:mixte}(a) shows the parallel electron temperature $T_{e\parallel}$ in Run C with a strong guide field. Sample in-plane projections of 3D magnetic field lines are plotted in blue. The bottom blue line is approximately within the separatrix layer and delineates the magnetospheric inflow from the reconnection exhaust. The density and parallel temperature profiles across the current sheet are plotted in Fig.~\ref{fig:mixte}(b) [along the cut marked in green in Fig.~\ref{fig:mixte}(a)]. As observed in the MMS data survey \cite{wang:2017}, $T_{e\parallel}$ drops within the inflow region even as the total electron density continues to increase. This occurs several times farther from the magnetic separatrix than the electron Larmor radius $\rho_e$, plotted Fig.~\ref{fig:mixte}(b). This indicates that the mixing results not only from finite electron orbit effects, which are limited to $\rho_e$ length scales and also occur in 2D \cite{hesse:2014,egedal:2016,zenitani:2017}. To demonstrate that this region indeed contains magnetosheath electrons mixed into the magnetosphere inflow, reduced electron velocity distributions in the $v_\parallel-v_\perp$ (directions with respect to local magnetic field) are plotted in Fig.\ref{fig:mixte}(c). Because the numerical particles are tagged as originating on the magnetosphere or magnetosheath sides of the simulation, we are able to generate separate plots for the magnetosheath and magnetosphere electrons. At point $I$ [marked in Figs.\ref{fig:mixte}(a-b)] roughly $4\rho_e$ into the magnetosphere inflow, a large population of relatively low-energy magnetosheath electrons is present. These electrons account for the increased density and the lower electron temperature. The magnetosheath population becomes smaller deeper into the magnetosphere inflow, as seen for points $II$ and $III$.

\section{Ohm's Law and Measuring Anomalous Resistivity}
\label{sec:ohm}

The anomalous particle transport induced by lower-hybrid fluctuations is associated with a corresponding anomalous resistivity \cite{davidson:1975}. Historically, there was strong interest in anomalous resistivity because observed reconnection rates are far faster than Sweet-Parker estimates in a plasma with classical resistivity \cite{parker:1957}. It was proposed, for example, that LHDI could generate anomalous resistivity and enhance reconnection rates in Earth's geomagnetic tail \cite{huba:1977}. Indeed, for plasma and magnetic field parameters typical of magnetospheric reconnection sites, order of magnitude estimates suggest that anomalous resistivity from lower-hybrid range fluctuations could be important for breaking the frozen-in condition near the X-line \cite{yoon:2007}. It has become clear, however, that reconnection can be fast in thin current sheets when kinetic or multi-fluid effects are taken into account \cite{kleva:1995,ma:1996,biskamp:1997,kuznetsova:2001,birn:2001,daughton:2006,cassak:2007prl}. In collisionless space plasmas, laminar inertia and pressure tensor terms from 2D models \cite{hesse:2014,hesse:2016} are sufficient to yield the observed fast reconnection rates. Of course, in naturally formed 3D systems, anomalous dissipation may still be produced. To be a dominant factor in governing the reconnection rate, however, the anomalous dissipation would have to significantly broaden the electron layer to reduce the laminar kinetic contributions.

Quantifying the importance of the anomalous contributions compared to essentially 2D kinetic effects has proven challenging. For example, the simple theoretical estimates rely on a number of approximations, notably assuming purely electrostatic fluctuations in a uniform equilibrium. Near the X-line of reconnecting current sheets, electromagnetic instabilities or corrections for high plasma $\beta$ \cite{davidson:1977,ozaki:1996,tanaka:1981,daughton:2003,kulsrud:2005} and stabilization by magnetic shear \cite{krall:1977} should be taken into account. Furthermore, the growth of any given instability must compete with the fast convection time for a fluid element transiting the diffusion region \cite{roytershteyn:2012}. Because of these and other complications, evaluating the contribution of kinetic instabilities within realistic reconnection geometries has relied heavily upon on numerical simulations for both the linear \cite{daughton:2003} and the non-linear regimes \cite{brackbill:1984,pritchett:1996,lapenta:2002,daughton:2004,silin:2005,che:2011,roytershteyn:2012,price:2016,le:2017}. Meanwhile, in laboratory reconnection experiments, lower-hybrid range fluctuations were observed within reconnecting current sheets \cite{carter:2001,ji:2004,fox:2010}. Both 3D kinetic modeling \cite{roytershteyn:2013} and subsequent experiments \cite{dorfman:2013} found that the fluctuations were too weak to substantially modify the reconnection physics.

\subsection{Spatial $y$-Averaging}

As a first method to attempt to quantify the turbulent or anomalous contributions to the reconnection electric field, we use $y$-averaging \cite{che:2011,roytershteyn:2012,price:2016,che:2017,le:2017} as was used in Section~\ref{sec:parttransport}. We start with the electron momentum balance, or Ohm's Law, equation,
\begin{equation}
ne({\bf{E}}+{\bf{u}}\times{\bf{B}}) = m\left[\frac{\partial (n\bf{u})}{\partial t}+\nabla\cdot (n{\bf{u}}{\bf{u}})\right] + \nabla\cdot \mathbb{P}_e,
\label{eq:ohm}\end{equation}
where $\mathbb{P}_e$ is the electron pressure tensor. Then the y-averaged momentum balance equation yields:
\begin{equation}
\yavg{n} e (\yavg{\bf{E}}+\yavg{{\bf{u}}}\times\yavg{{\bf{B}}}) = m\left(\frac{\partial (\yavg{n\bf{u}})}{\partial t}+\nabla\cdot[(\yavg{n{\bf{u}}})(\yavg{{\bf{u}}})]\right) + \nabla\cdot \yavg{\mathbb{P}_e} - e\yavg{\delta n \delta{\bf{E}}} - e \yavg{\delta(n{\bf{u}})\times\delta{\bf{B}}} + m\nabla\cdot[\yavg{\delta(n{\bf{u}})\delta{\bf{u}}}],
\label{eq:avgohm}\end{equation}
where we have chosen to retain the particle flux $n{\bf{u}}$ as a single quantity when decomposing products of terms. The anomalous terms are those that contain correlated fluctuations, and $e\yavg{\delta n \delta{\bf{E}}}$ is commonly referred to as an anomalous resistivity. The remaining two terms have been combined and called anomalous viscosity \cite{che:2011,price:2016}. To differentiate between them here, we will refer to $e \yavg{\delta(n{\bf{u}})\times\delta{\bf{B}}}$ as the anomalous Lorentz force and $m\nabla\cdot[\yavg{{\delta(n{\bf{u}})\delta{\bf{u}}}}]$ as the Reynold's stress contribution.

\begin{figure}
\includegraphics[width = 12.0cm]{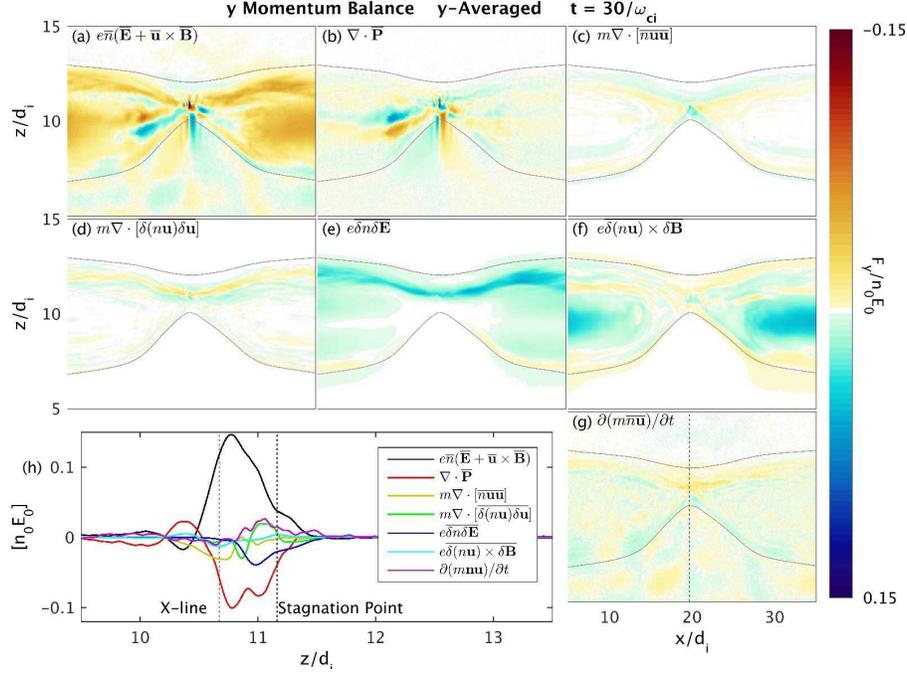}
\caption{Terms in the y-averaged Ohm's law contributing to electron momentum balance from Run A with weak guide field. Anomalous resistivity $\propto\yavg{\delta n\delta{\bf{E}}}$ contributes appreciably to the non-ideal electric field near the stagnation point at the simulation time ($t=30/\omega_{ci}$) shown.   However, this is likely a transient effect since at later times the relative importance is diminished. \label{fig:yavg-a}}
\end{figure}

\begin{figure}
\includegraphics[width = 12.0cm]{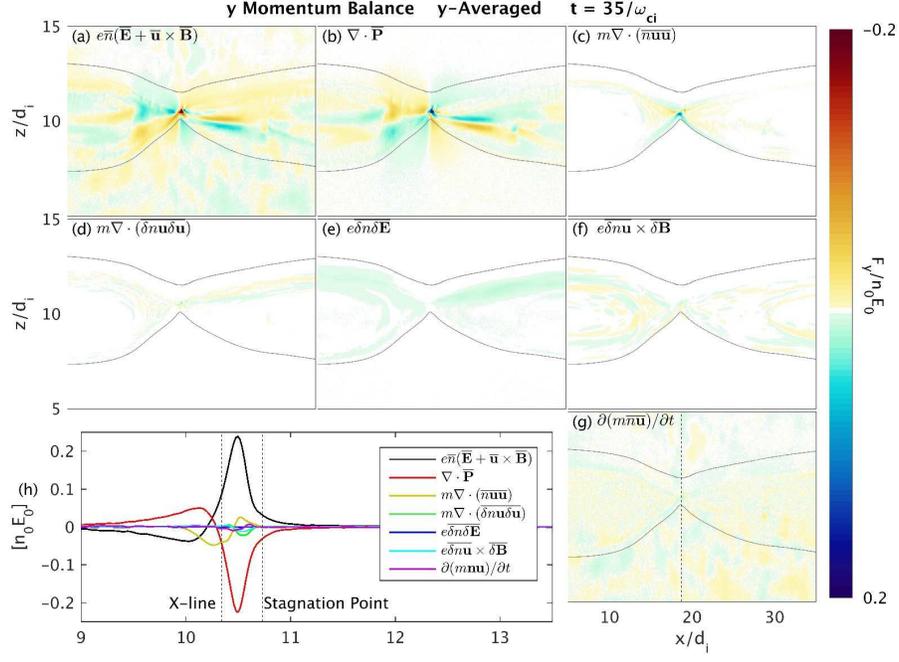}
\caption{Terms in the y-averaged Ohm's law contributing to electron momentum balance from Run B with $B_g/B_0\sim0.4$. The anomalous terms are negligible. \label{fig:yavg-b}}
\end{figure}

\begin{figure}
\includegraphics[width = 12.0cm]{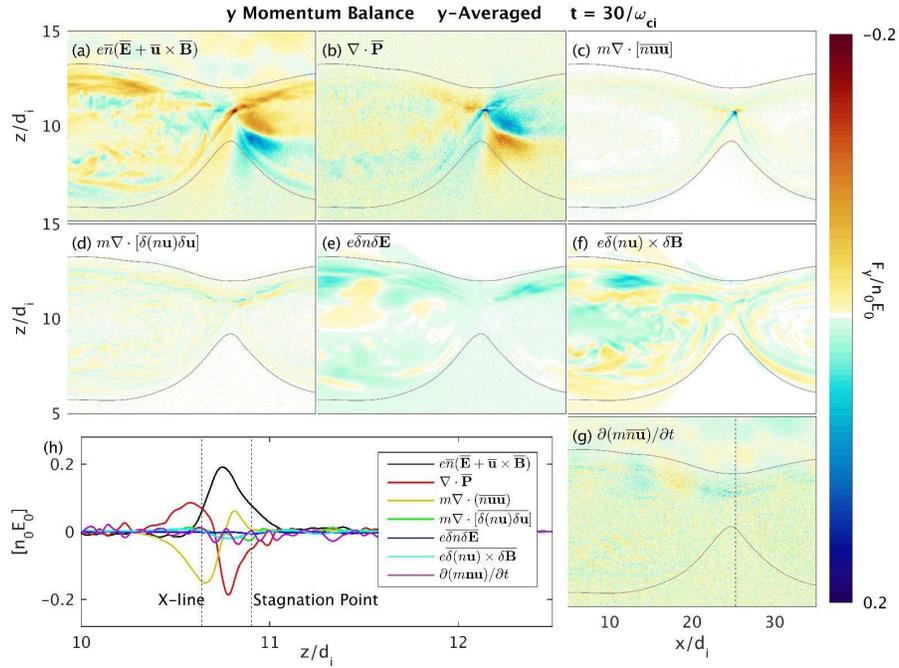}
\caption{Terms in the y-averaged Ohm's law contributing to electron momentum balance from Run C with order-unity guide field. The anomalous terms are negligible. \label{fig:yavg-c}}
\end{figure}

The various contributions to the $y$-averaged out-of-plane electron momentum balance equation from each 3D simulations are plotted for Run A in Figs.~\ref{fig:yavg-a}(a-g), Run B in Figs.~\ref{fig:yavg-b}(a-g), Run C in Figs.~\ref{fig:yavg-c}(a-g). Panel (h) of each figure shows a line out of the terms along a cut through the X-line across the current sheet [indicated in each panel (g)]. In asymmetric reconnection, the X-line and the electron flow stagnation may be spatially separated \cite{cassak:2007}. For 2D systems without a guide field, the reconnection electric field in the region between the X-line and stagnation point was found to be supported by electron inertia (especially near X-line) and the divergence of the electron pressure tensor \cite{hesse:2014}. In Runs B and C with moderate and stronger guide fields, we find that the 2D-like inertial and pressure terms continue to dominate in the Ohm's law. The anomalous contributions are small throughout these simulations.

\begin{figure}
\includegraphics[width = 12.0cm]{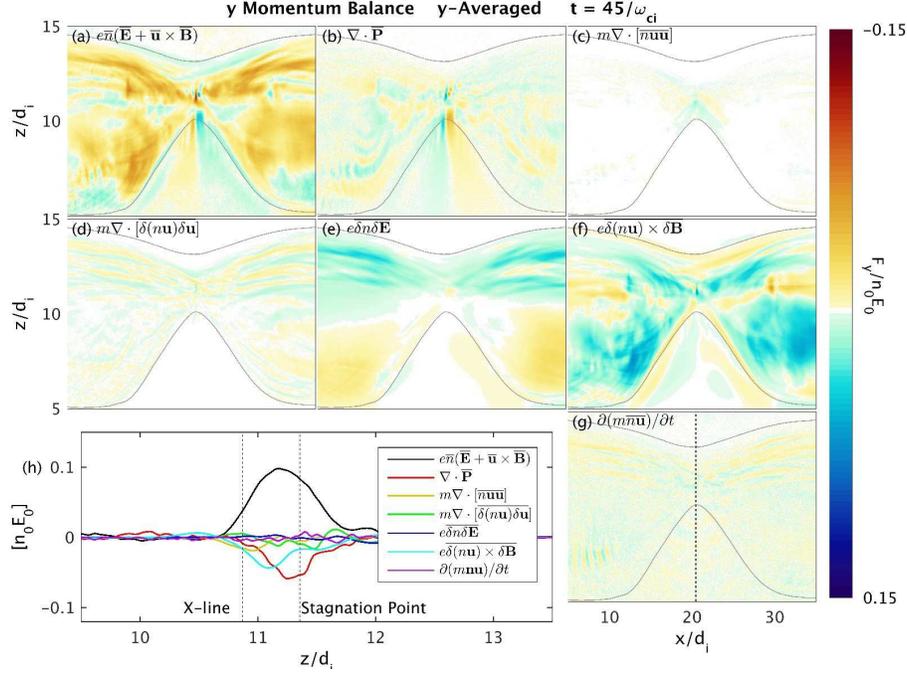}
\caption{Terms in the y-averaged Ohm's law contributing to electron momentum balance from Run A with weak guide field at late time in the simulation. At this stage of the simulation, the anomalous Lorentz force $\propto\yavg{\delta(n{\bf{u}})\times\delta{\bf{B}}}$ piece of the anomalous viscosity is large. \label{fig:yavg-a-45}}
\end{figure}

In Run A with a weak guide field, $y$-averaging does produce anomalous contributions to Ohm's Law.  In Fig.~\ref{fig:yavg-a}(h), for example, it is seen that the anomalous resistivity $\propto\yavg{\delta n \delta E_y}$ is as large as the electron inertia, particularly near the stagnation point. However, as noted in earlier papers \cite{le:2017,price:2017}, this anomalous resistivity is transient, and it subsides after the initial burst of LHDI becomes weaker. This implies that the observed anomalous resistivity is driven by the sharp gradients set up by the initial conditions containing a thin sheet and magnetic perturbation.  Figure~\ref{fig:yavg-a-45} shows the contributions to the $y$-averaged Ohm's Law at a later time of $t\omega_{ci}=45$. In the cuts through the X-line in Fig.~\ref{fig:yavg-a-45}(h), the anomalous resistivity is negligible. On the other hand, the anomalous Lorentz $\propto\yavg{\delta(n{\bf{u}})\times\delta{\bf{B}}}$ term is large, and even dominant, over a broad region between the X-line and the stagnation point, in agreement with previous simulations \cite{price:2017}. We attribute this to the development of the electromagnetic LHDI \cite{daughton:2003} that results in a kinking of the current sheet over a longer time scale. Recall, however, that the overall reconnection rate is similar for this run in 2D and in 3D.

One complication of the $y$-averaged Ohm's Law is that its interpretation assumes that, on average, the system remains translationally invariant in the $y$-direction. The current sheet, however, may be unstable to a kink instability [most apparent in the weak guide field Run A in Fig.~\ref{fig:runs}] with a wavelength several times larger than the electrostatic LHDI fluctuations. The observed wavelength of the kink instability observed in the present simulation ($k\sim2\pi/L_y\sim0.6/d_{i0}$) is consistent with the longer wavelength electromagnetic LHDI \cite{daughton:2003,daughton:2004} predicted to be most unstable for $k\sqrt{\rho_e\rho_i}\sim1$. This mode has been studied using both Vlasov theory and kinetic simulations up to the physical mass ratio for hydrogen ($m_i/m_e=1836$) \cite{daughton:2003}, confirming its weak wavelength dependence $\propto(m_i/m_e)^{1/4}$ on the mass ratio. It has been observed in spacecraft observations \cite{zhou:2009} and in laboratory experiments \cite{roytershteyn:2013}. While the mode survives for weak guide field \cite{daughton:2003}, the strong guide field limit has not been studied in detail. Nevertheless, the line-bending term associated with the guide field is expected to be stabilizing, although we cannot rule out the possibility that the kinking is suppressed in the guide field runs because the finite domain is too short to support the most unstable mode.

\begin{figure}
\includegraphics[width = 12.0cm]{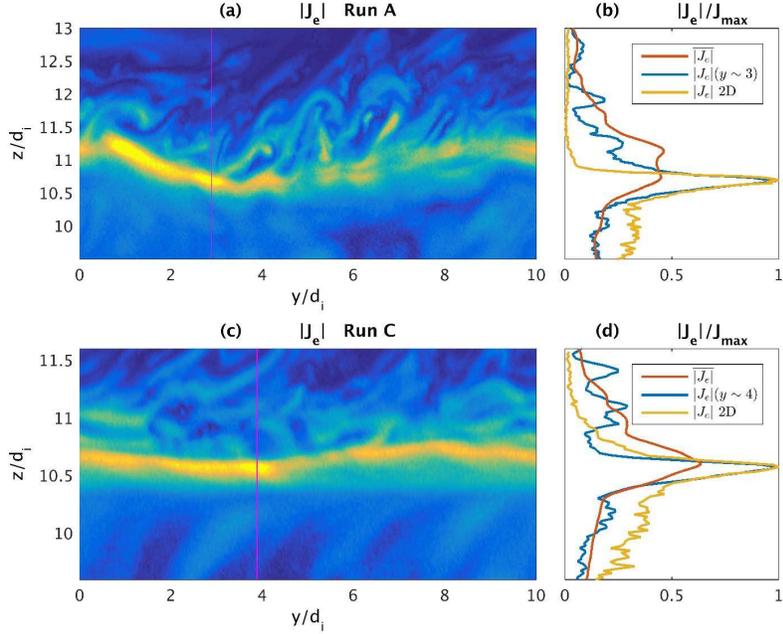}
\caption{The electron current density $|J_e|$ in the $y$ (principal current flow direction) and $z$ (normal to the current sheet) plane at the $x$ location of the X-line from (a) Run A with weak guide field and (c) Run C with strong guide field. (b,d) Cuts through the normal $z$ direction of the $y$-averaged current density $\overline{|J_e|}$, of the current density through the cuts indicted in (a,c), and from the corresponding 2D simulations. The current sheet width is locally similar in 2D and 3D, while the $y$-averaged profile appears broader because of the longer wavelength kink of the current layer.\label{fig:Je}}
\end{figure}

Even a moderate kinking of the current sheet entails variations in the $y$-direction. See, for example, Fig.~\ref{fig:Je}(c), which shows a cut in the $y-z$ plane of the electron current density at the approximate X-line in Run C with a strong guide field. The strong guide field of Run C tends to stabilize kink instabilities, and the kinking is considerably weaker than in the low guide field case of Run A [Fig.~\ref{fig:Je}(a)]. Even so, the small kink of the current sheet creates a broader current profile when the current density is $y$-averaged [see Fig.~\ref{fig:Je}(b)]. Locally, however, the current density remains similarly peaked in 2D and 3D with a full width at half max of $\sim 2d_{e0}$ in each of the 3D runs, suggesting that the underlying dissipation physics is the same.

\begin{figure}
\includegraphics[width = 12.0cm]{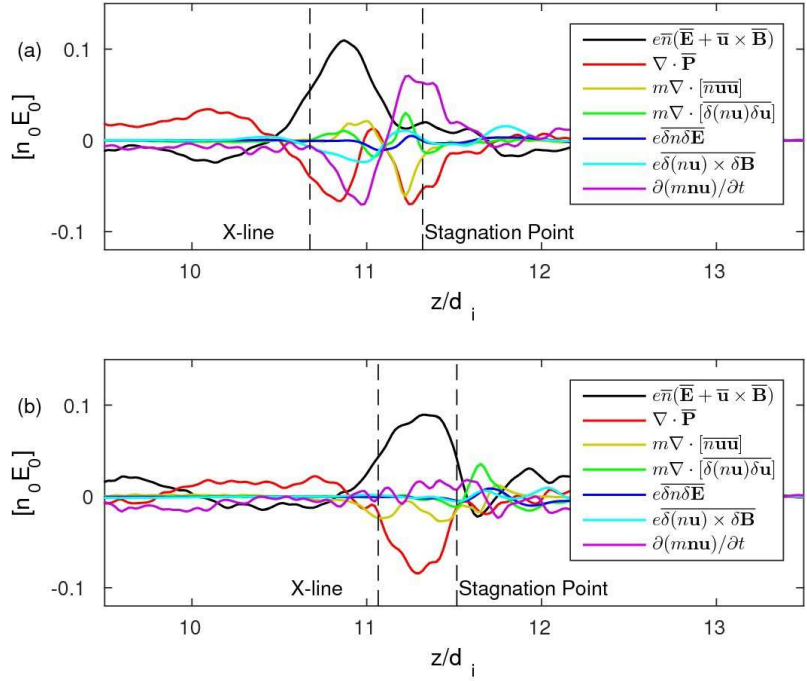}
\caption{Cuts across the current sheet through the X-line of the $y$-averaged Ohm's law from Run A at $t=45/\omega_{ci}$, with the average taken over a subset of the simulation domain in the range (a) $y\in[0,2.5]d_{i0}$ and (b) $y\in[5,7.5]d_{i0}$. The anomalous terms are not consistent with averaging over the whole domain in Fig.~\ref{fig:yavg-a-45}(h), suggesting that they do not result from the short wavelength fluctuations. \label{fig:ycuts}}
\end{figure}

To highlight the effect of the longer wavelength kinking on the y-averaged Ohm's law, we plot cuts of the electron momentum balance equation in Fig.~\ref{fig:ycuts} where the $y$-average is taken over two different subsets of the domain of length $2.5d_{i0}$ (one quarter of the simulation). This range covers $\sim3$ wavelengths of the electrostatic LHDI fluctuations, and so if the short wavelength waves contributed consistently to anomalous dissipation over the length of the simulation box, the contributions to the Ohm's law should be similar for the two cases. On the contrary, the first segment for $y\in[0,2.5]d_{i0}$ [Fig.~\ref{fig:ycuts}(a)] is from a region where the current is kinked and misaligned with the $y$ direction [see Fig.~\ref{fig:Je}(a)]. Here, there is a large contribution from the time derivative term. For $y\in[5,7.5]d_{i0}$ [Fig.~\ref{fig:ycuts}(a)] on the other hand, the phase of the kink leaves the current sheet locally nearly aligned with $y$-direction, and the Ohm's law looks essentially 2D, with very small contributions from anomalous or time-varying terms. This strongly suggests that the anomalous viscosity that appears in Fig.~\ref{fig:yavg-a-45} is thus not caused by shorter wavelength contributions that add up consistently over the full length of the domain, but it is rather a result of the $m=1$ longer wavelength kinking of the current sheet.  

\subsection{Temporal $t$-Averaging}
As an alternative to $y$-averaging, which can introduce spurious broadening of spatial profiles, we evaluate here time-averaging. We define the $t$-average of any quantity $Q$ as
\begin{equation}
\tavg{Q}(x,y,z,t_0) = \frac{1}{\Delta t}\int_{t_0}^{t_0+\Delta t} Q(x,y,z,t) dt
\end{equation}
For Run C, we averaged over intervals of 200 numerical time steps, so that $\Delta t\sim5.6/\omega_{ce0}\sim0.65/\omega_{lh}$, where $\omega_{lh}$ is the lower-hybrid frequency evaluated within the turbulent magnetospheric boundary layer. Unfortunately, for Run C the averaging interval therefore did not cover a full period of the lower-hybrid fluctuations. Because of the large data storage requirements of these 3D simulations, the time-averaging must be performed in-line. Changing the time-averaging interval would thus require re-running the simulation, which is not presently feasible. For Run B, on the other hand, the time-averaging interval was set to 2000 numerical total time steps with every tenth step included in the average. In this way, the averaging interval covered one full cycle of the lower-hybrid range fluctuations with $\delta t\sim6.3/\omega_{lh}\sim 2\pi/\omega_{lh}$. (The time-averaging diagnostic was unavailable for Run A.)

\begin{figure}
\includegraphics[width = 12.0cm]{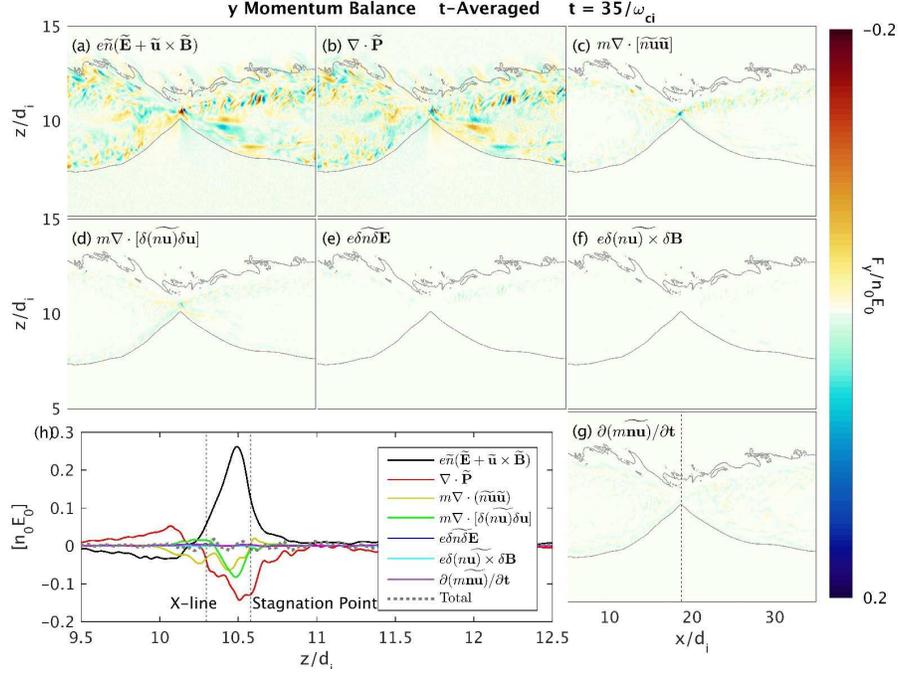}
\caption{Terms in the $t$-averaged Ohm's law contributing to electron momentum balance from Run B with $B_g/B_0\sim0.4$. A contribution from the Reynold's stress $\propto\nabla\cdot[m\tavg{\delta(n{\bf{u}})\delta{\bf{u}}}]$ appears between the X-line and the stagnation point. \label{fig:tavgb}}
\end{figure}

The contributions to the $t$-averaged Ohm's Law from Run B are plotted in Fig.~\ref{fig:tavgb}. The dashed gray curve marked "Total" is the sum of the contributions, and small deviations from zero indicate numerical noise. We find in this case that the Reynold's stress $\propto\nabla\cdot[m\tavg{\delta(n{\bf{u}})\delta{\bf{u}}}]$ takes on values comparable to the reconnection electric field. The values oscillate, however, and they take opposite signs for different choices of the 2D $x-z$ plane in the simulation volume. The other contributions are similar to 2D, with inertia and the pressure tensor both contributing to the non-ideal electric field. 

\begin{figure}
\includegraphics[width = 12.0cm]{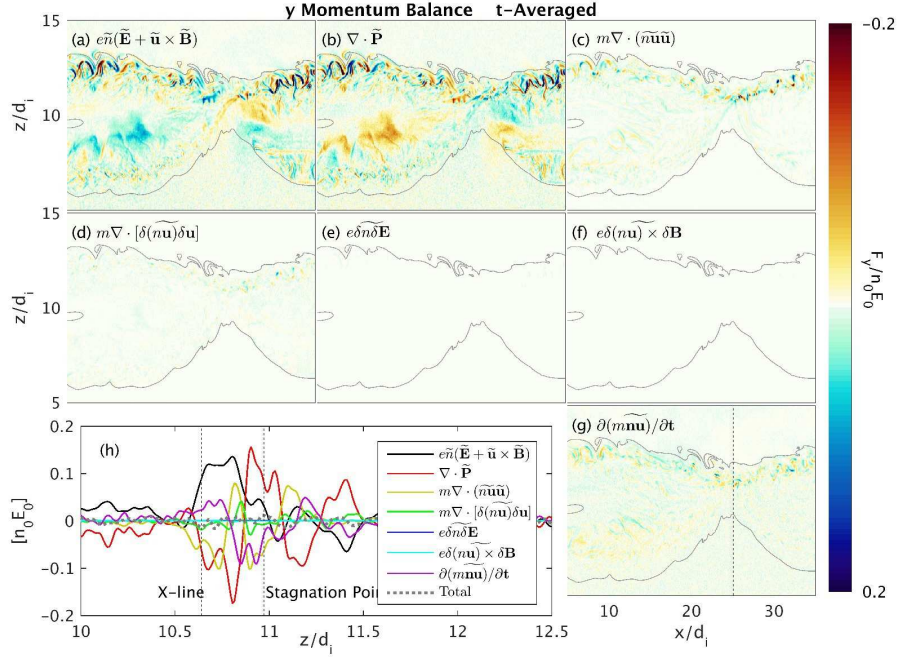}
\caption{Terms in the $t$-averaged Ohm's law contributing to electron momentum balance from Run C with order-unity guide field. The time-averaging interval was here not long enough to cover the entire lower-hybrid period, and significant contributions from drift fluctuations along the magnetosphere separatrix remain in the time-derivative (g) term. "Anomalous" terms (d-f) on this scale were all negligible. \label{fig:tavgc}}
\end{figure}

In Fig.~\ref{fig:tavgc}, the various contributions to the $t$-averaged Ohm's Law are plotted from Run C with $B_g/B_0\sim1$. For Run C, the anomalous contributions [Figs.~\ref{fig:tavgc}(d-f) are negligible. However, because the interval $\Delta t$ was shorter than the full lower-hybrid period for Run C, fluctuations on the lower-hybrid time scale remain in the "mean" quantities. The mean terms in Figs.~\ref{fig:tavgc}(a-c) therefore show considerable fine-scale structure along the magnetospheric separatrix. These small spikes of non-ideal electric field are balanced by averaged time derivative $\partial/\partial t$ terms. The short time averaging interval thus keeps characteristics of the local, un-averaged Ohm's law of Eq.~\ref{eq:ohm}. See the Appendix for a comparison to the local, unaveraged Ohm's law for Run C.

Unfortunately, we did not test the effect of varying the $t$-averaging interval when these large calculations were performed. The difficulty in interpreting the results of the averaging, however, do highlight the complications involved in comparing numerical data to spacecraft observations, which necessarily average over the finite cadence of the measuring instruments \cite{price:2017,torbert:2016,torbert:2017}. MMS electromagnetic field data is collected at a high enough cadence that lower-hybrid fluctuations in the 30-40 Hz range are easily resolved. This time scale is somewhat under-resolved, however, by the 30 ms burst-mode FPI 3D electron distribution data.

\subsection{Temporal and Field Line $t\ell$-Averaging}

The time-averaging procedure described in the previous section gives a local measure of anomalous dissipation that is less sensitive to potentially misleading kinking of the current layer out of the initial symmetry direction.  However, the relative importance of various terms may depend upon the chosen location (x-z plane) chosen within the 3D volume.  Ideally, we would like to characterize the importance of each physical term in Ohm's law on a field-line basis, and then correlate this understanding with special field lines that pass through the diffusion region or QSL. Furthermore, the simulations do not remain translationally invariant in the out-of-plane $y$ direction, even on average. Thus, while the electric field $E_y$ is the local reconnection electric field at the X-line in 2D systems, other components of the electric field may play this role in a fully 3D system. We seek a diagnostic that does not depend on an initial symmetry direction of the simulation and that could be applied in more realistic 3D systems. A fairly general definition of magnetic reconnection in 3D is based on the electric field $E_\parallel$ parallel to the magnetic field \cite{schindler:1988}. Under certain conditions, the reconnection rate may be defined as the maximum of the magnetic field line-integrated parallel electric field $E_\parallel$ \cite{hesse:1993,liu:2015,sauppe:2018}. While the assumptions of this theory are not formally satisfied in the periodic slab geometry of our simulations, the peak line-integrated $E_\parallel$ has still been found to occur on the field line that passes through the approximate X-line and has a value that agrees with other measures of the reconnection rate. 

In light of the above considerations, we consider the component of the electron momentum balance equation in the direction of the magnetic field integrated along field lines. We define the following $t\ell$-average over time and magnetic field lines:  
\begin{equation}
\flavg{{\tavg{\bf{V}}}}(x_0,y_0,z_0,t_0) = \frac{1}{L}\int_{{\bf{x}}_0}^{\bf{x_f}} {\bf{\tavg{b}}}\cdot{\bf{\tavg{V}}}(x,y,z,t_0) d\ell
\end{equation}
where ${\bf{\tavg{b}}} = {\bf{\tavg{B}}}/|\tavg{B}|$ is the unit vector in the direction of the time-averaged magnetic field, and the integral is taken from each initial point $\bf{x}_0$ within the sample plane to a final point $\bf{x}_f$ a distance $L=5d_{i0}$ away along the magnetic field. To test this method, we choose to seed our field line integrator with initial points $\bf{x}_0$ that populate a plane in the $x$ and $z$ directions. In the Appendix, we verify that using the unit vector and fields lines of the time-averaged magnetic field $\tavg{\bf{B}}$ does not significantly affect the conclusions of our calculations.

\begin{figure}
\includegraphics[width = 12.0cm]{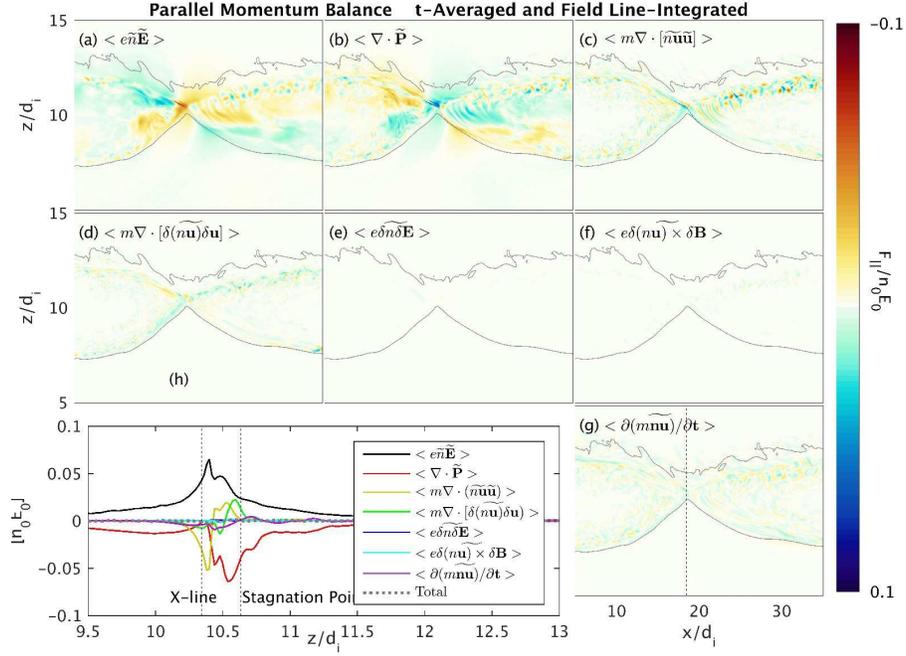}
\caption{Terms in the t-averaged and field-line integrated parallel Ohm's law from Run B with $B_g/B_0\sim0.4$. The anomalous Reynold's stress contribution(green curve) is found on average to be opposite that required to balance the global reconnection electric field. \label{fig:tlavgb}}
\end{figure}

\begin{figure}
\includegraphics[width = 12.0cm]{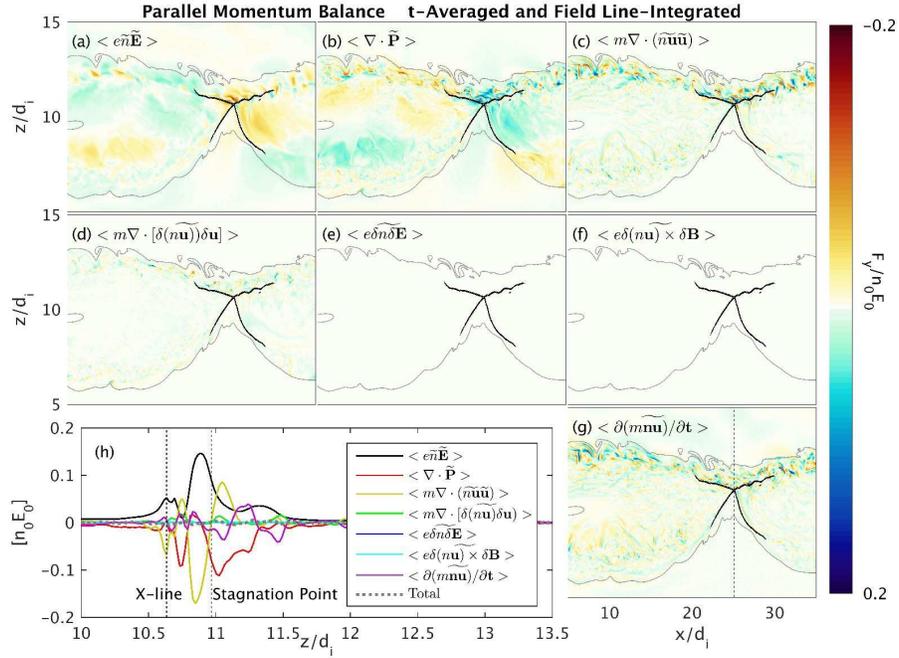}
\caption{Terms in the t-averaged and field line-integrated parallel Ohm's law from Run C with order-unity guide field. Anomalous terms are small. \label{fig:tlavgc}}
\end{figure}

The various terms in the $t\ell$-averaged Ohm's Law are plotted in Fig.~\ref{fig:tlavgb} from Run B with $B_g/B_0\sim0.4$ and in Fig.~\ref{fig:tlavgc} from Run C with $B_g/B_0\sim1$. While we plot these values in panels (a-g) of each figure at the initial seed points in a plane, each term is really a field-line average that must be associated with an entire magnetic field line. Thus, in the line cuts in panels (h) in each plot, the contributions labeled "X-line" and "Stagnation Point" are not local values assigned to those points, but rather they are average contributions along the magnetic field lines that pass through those special points in our selected seed plane. For Run C in Fig.~\ref{fig:tlavgc}, the anomalous terms are all negligible compared to mean 2D-like terms. For Run B in Fig.~\ref{fig:tlavgb}, there is a contribution from the anomalous Reynold's stress that is comparable to the reconnection electric field. As noted above, this anomalous term oscillates, and its sign depends on the which $x-z$ plane in the 3D volume we select. For this particular plane of seed points, its average value on the field line through the X-line is opposite to the spatially local value found in Fig.~\ref{fig:tavgb}. Nevertheless, the picture that emerges is consistent with 2D simulations \cite{hesse:2016,egedal:2018}: inertia and the divergence of the pressure tensor dominate between the X-line and the stagnation point. 

\section{Summary Discussion}

Three fully kinetic 3D simulations of asymmetric magnetic reconnection were performed with plasma parameters based on MMS diffusion region encounters. A common feature of all three simulations was the development of lower-hybrid range fluctuations, driven by diamagnetic drifts along the steep density gradient separating magnetosheath and magnetosphere plasmas. 

Three methods of measuring the contributions to the Ohm's law of the drift fluctuations were considered. In Table~\ref{tab:ohm}, we summarize the results for each run employing the $y$-, $t$-, and $t\ell$-averaging measures. Overall, the picture of electron momentum balance is similar to 2D, where inertia and the divergence of the electron pressure tensor may both be important in the region between the X-line and the flow stagnation point \cite{hesse:2016}. In Run A, a transient burst of anomalous resistivity is found early in the simulation, though this is most likely related to the choice of initial equilibrium current sheet. If spatial $y$-averaging is employed, the anomalous Lorentz contribution to the viscosity appears important at later time in agreement with Ref.~\cite{price:2017}. However, spatial averaging may be problematic when the current sheet undergoes kinking because this will create a broadened averaged profile, while the current sheet may remain thin locally. Indeed, the anomalous viscosity was not consistently observed when the average was taken over shorter spatial intervals in the $y$-direction, implying it is caused by the longer wavelength kinking of the current sheet rather than the shorter wavelength electrostatic LHDI. The only other evidence of a non-negligible turbulent contribution to the reconnection electric field comes from a Reynold's stress contribution to anomalous viscosity in the intermediate guide field Run B when temporal averaging is used to define fluctuations. This contribution, however, takes both signs depending on the $x-z$ cut plane, and it does not consistently contribute in either direction over the full volume of the simulation. Thus, we did not find any consistent anomalous contributions to the reconnection electric field at the X-line from the short wavelength LHDI. Furthermore, under conditions typical of the magnetopause for the range of guide fields considered, the global reconnection rates observed in our 3D simulations were very close to 2D simulations 

\begin{table*}
\caption{Contributions to the averaged Ohm's Law in each run. Fluctuating contributions are listed in $\bf{bold}$ text. Anomalous Resistivity is abbreviated as $\bf{AR}$, the Anomalous Lorentz contribution is listed as $\bf{AL}$, and $\bf{RS}$ refers to the Reynold's Stress term. For Run A, the $t$-averaging diagnostic was unavailable. While the non-ideal field typically peaks between the X-line and stagnation points, we indicate which terms are largest at those two points. \label{tab:ohm}}
\begin{tabular*}{15cm}{|c|p{3cm}|p{3cm}|p{3cm}|p{3cm}|}
    \hline\hline
    Run 		& {} 				& $y$-Average 						& $t$-Average 		& $t\ell$-Average \\ \hline\hline
    A 			& X-line			& Inertia\newline divP (early)\newline \bf{AL} (late)	& $\times$ 		& $\times$ \\ \cline{2-5}
    $B_g/B_0\sim 0.1$   &Stagnation Point 			& divP\newline \bf{AR} (early) \newline AL (late)	& $\times$ 		& $\times$ \\ \hline
    B  			& X-line 			& Inertia\newline divP 					&Inertia\newline divP  	& Inertia \\ \cline{2-5}
    $B_g/B_0\sim 0.4$   &Stagnation Point			& divP 							& divP\newline (\bf{RS})& divP \newline (\bf{RS})   \\ \hline
    C  			&X-line	& Inertia						&Inertia\newline divP 	& Inertia\\ \cline{2-5}
    $B_g/B_0\sim 1$     & Stagnation Point		& divP 							&Inertia		& divP \newline Inertia \\\hline\hline
\end{tabular*}
\end{table*}

In the classic picture, anomalous dissipation arises from small amplitude fluctuations on small spatial scales and fast temporal scales that are well-separated from the global dynamics. These assumptions are not very well satisfied in our reconnection simulations. This muddies the interpretation of the anomalous dissipation terms, which depend on the choice of averaging scales and procedure. For comparison to spacecraft data, too, it is unclear how best to quantify or identify anomalous dissipation.  This makes identifying the reconnection rate difficult because the local fluctuating electric fields may be orders of magnitude larger than the global reconnection electric field. Despite the difficulty of this measurement, analysis of the Ohm's law in MMS encounters with diffusion regions have still found that high-frequency or short-wavelength fluctuations may have contributions as large as the resolved terms \cite{torbert:2016,torbert:2017} in the non-ideal electric field, even when using electron moments inferred at a 7.5 ms cadence \cite{rager:2018} . 

While the overall reconnection rates were similar in 2D and 3D, the drift fluctuations along the magnetospheric separatrix produced particle transport across the magnetic field that relaxed the density gradient across the magnetopause \cite{price:2017}. The level of particle transport was found to be in reasonable agreement with estimates from quasi-linear theory as well as diffusion coefficients inferred from MMS observations \cite{graham:2017}. A signature of the enhanced particle transport is the presence of cold beams of magnetosheath electrons within the magnetospheric inflow, which were observed by MMS \cite{burch:2016grl} and in our 3D kinetic simulations. In each 3D simulation, there was also enhanced heating of the electrons (compared to 2D simulations) in the parallel direction within the mix layer. In 2D simulations, which limit cross-field particle transport, the inflow electron heating was well-described by CGL-like equations of state \cite{egedal:2011pop,le:2009,le:2017}. In our 3D simulations and MMS data, however, the electron temperature may decrease even as the density decreases \cite{wang:2017,graham:2017}. A complete heating model must account for the mixed magnetosheath electrons as well as the precise heating mechanisms in the mix layer, which are left for future work.  

\begin{acknowledgments}
A.L. received support from the LDRD office at LANL and acknowledges NASA project NNX14AL38G/NNH17AE36I. W.D.’s contributions were supported by the Basic Plasma Science Program from the DOE Office of Fusion Energy Sciences, and by NASA grant NNH13AW51I. Y.-H. L. is supported by NASA grant NNX16AG75G and the MMS mission. Simulations were performed on Trinity at LANL, Blue Waters at the NCSA through project ACI1640768, Pleiades provided by NASA's HEC Program, and LANL Institutional Computing resources. 
\end{acknowledgments}

\appendix
\section{Using $\tavg{\bf{b}}$}
\begin{figure}
\includegraphics[width = 12.0cm]{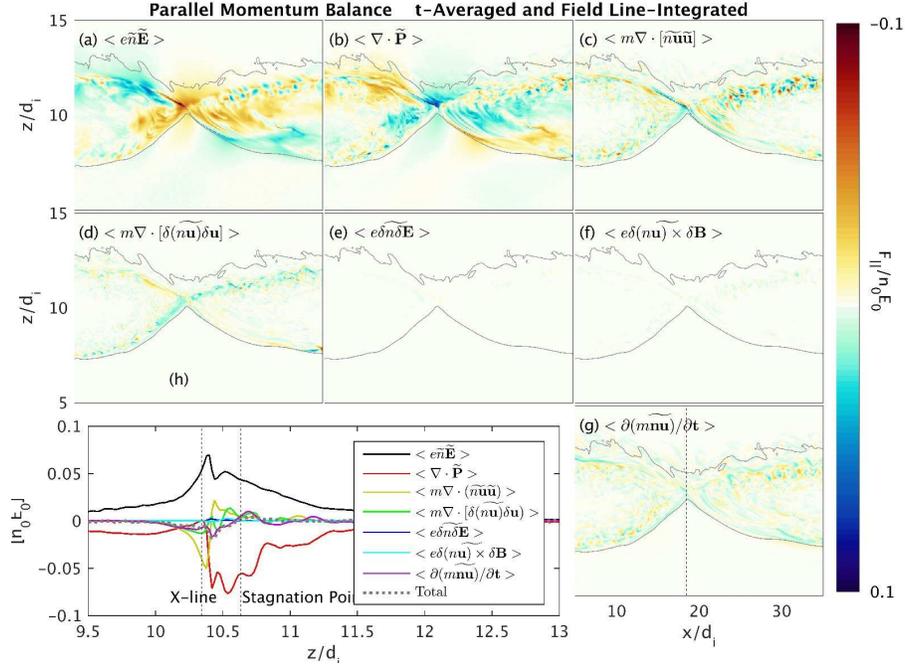}
\caption{Terms in the t-averaged and field-line integrated parallel Ohm's law from Run B with $B_g/B_0\sim0.4$ using the magnetic field ${\bf{B}}(t_0)$ at the beginning of the $t$-averaging interval. The general balance of terms is similar to using the $t$-averaged field $\tavg{\bf{B}}$ [see Fig.~\ref{fig:tlavgb}], although the anomalous Reynold's stress term is smaller in this case. \label{fig:partlavgB-b}}
\end{figure}

In our definition of the $t\ell$-average over time and field lines, we integrate over the field lines defined by the time-averaged magnetic field $\tavg{\bf{B}}$ and we take the dot product of each term with the unit vector $\tavg{\bf{b}}$ in the direction of the time-averaged magnetic field. Of course, the magnetic field lines are only well-defined at each given instant $t$, and they cannot in general be tracked over time. Our definition also neglects terms that may arise from fluctuations in the direction of the magnetic field, $\delta\bf{b}$. To verify that these complications do not affect our main conclusions, we repeated the $t\ell$-averaged Ohm's Law calculation using the magnetic field at the initial instant of the $t$-averaging interval, which is to say for any quantity $Q$ we computed the average
\begin{equation}
\flavg{{\tavg{\bf{V}}}}(x_0,y_0,z_0,t_0) = \frac{1}{L}\int_{{\bf{x}}_0}^{\bf{x_f}} {\bf{b}}(t=t_0)\cdot{\bf{\tavg{V}}}(x,y,z,t_0) d\ell
\end{equation}
where now the integral is over field lines of ${\bf{B}}(t_0)$. The results are plotted from Run B, which used the longer time-averaging interval, in Fig.~\ref{fig:partlavgB-b}. The results are similar to the calculations using the time-averaged field $\tavg{\bf{B}}$ of Fig.~\ref{fig:tlavgb}, although the Reynold's stress term is somewhat smaller when using ${\bf{B}(t_0)}$. This confirms that the magnetic field does not vary too greatly over the course of our averaging time interval. For strongly electromagnetic fluctuations, however, a more general field line averaging method would have to be developed.

\section{Local Ohm's Law for Run C}
To obtain a clean picture of "anomalous" versus "mean" terms in the time-averaged treatment, we should use a time interval over which the average time derivative is negligible. In that case, the time-averaged reconnection region would appear quasi-stationary. For Run B, which had a time averaging interval covering the lower-hybrid period, this was the case. The time derivative terms in Fig.~\ref{fig:tavgb}(g) are small, and the averaged Ohm's law resembles a 2D picture with the "mean" inertial and pressure divergence terms balancing the non-ideal electric field. It turns out that the "anomalous" terms are also small.

\begin{figure}
\includegraphics[width = 12.0cm]{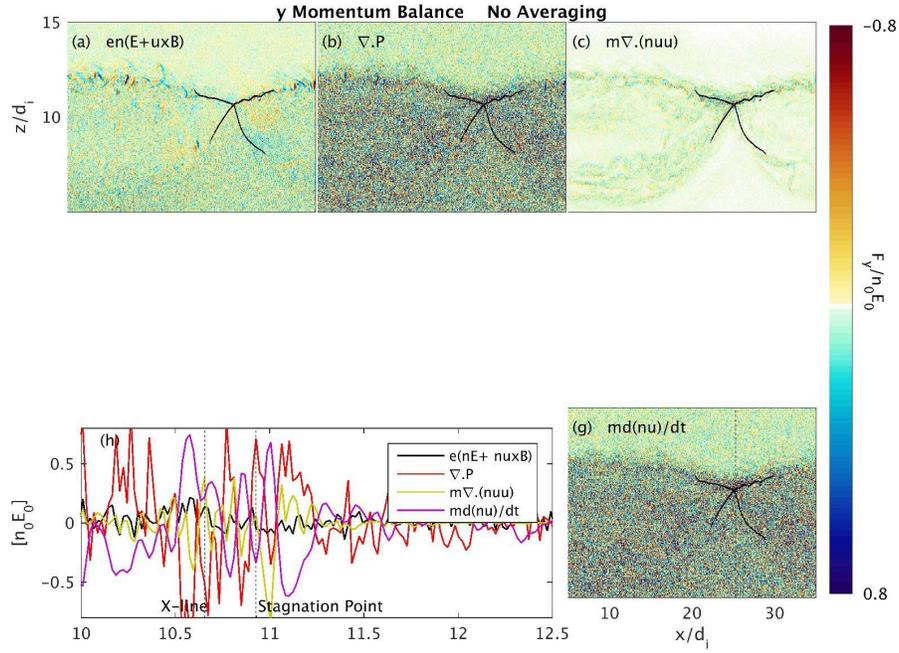}
\caption{Terms in the local Ohm's law (as in Eq.~\ref{eq:ohm}) from Run C with $B_g/B_0\sim1$. This figure can be compared to Fig.~\ref{fig:tavgc}, which shows the time-averaged Ohm's law using a relatively short averaging interval $\Delta t$ in the same $y$ cut plane. (For this case, there are no anomalous terms associated with averaging, and the middle panels have been left blank.) \label{fig:ymom-local}}
\end{figure}

As noted, the time-averaging interval for Run C is short, and it only covers a fraction ($\sim0.1$) of a lower-hybrid wave period. As a result, significant contributions to the non-ideal electric field caused by drift fluctuations remain in the averaged time derivative. The shorter interval is thus inadequate for separating the fluctuations from an average quasi-stationary state, muddling the separation of "mean" and "anomalous" terms. Indeed, for this short averaging interval, the Ohm's law resembles more closely the local Ohm's law of Eq.~\ref{eq:ohm}. To illustrate this, the un-averaged contributions to Ohm's law are plotted in Fig.~\ref{fig:ymom-local} from Run C in the same plane as in Fig.~\ref{fig:tavgc}. On comparing Figs.~\ref{fig:tavgc} and \ref{fig:ymom-local}, the most obvious effect of the time averaging is likely the smoothing out of grid-scale noise associated with the PIC method. In the un-averaged Ohm's law, we find large time derivative terms [Fig.~\ref{fig:ymom-local}(g)]. While these are smaller in the time-averaged picture in Fig.~\ref{fig:tavgc}, the interval was nevertheless too short to fully smooth them out.

\begin{thebibliography}{91}
\expandafter\ifx\csname natexlab\endcsname\relax\def\natexlab#1{#1}\fi
\expandafter\ifx\csname bibnamefont\endcsname\relax
  \def\bibnamefont#1{#1}\fi
\expandafter\ifx\csname bibfnamefont\endcsname\relax
  \def\bibfnamefont#1{#1}\fi
\expandafter\ifx\csname citenamefont\endcsname\relax
  \def\citenamefont#1{#1}\fi
\expandafter\ifx\csname url\endcsname\relax
  \def\url#1{\texttt{#1}}\fi
\expandafter\ifx\csname urlprefix\endcsname\relax\def\urlprefix{URL }\fi
\providecommand{\bibinfo}[2]{#2}
\providecommand{\eprint}[2][]{\url{#2}}

\bibitem[{\citenamefont{Priest and Forbes}(2002)}]{priest:2002}
\bibinfo{author}{\bibfnamefont{E.~R.} \bibnamefont{Priest}} \bibnamefont{and}
  \bibinfo{author}{\bibfnamefont{T.~G.} \bibnamefont{Forbes}},
  \bibinfo{journal}{Astronomy and Astrophysics Review}
  \textbf{\bibinfo{volume}{10}}, \bibinfo{pages}{313} (\bibinfo{year}{2002}),
  ISSN \bibinfo{issn}{0935-4956}.

\bibitem[{\citenamefont{Yamada et~al.}(2010)\citenamefont{Yamada, Kulsrud, and
  Ji}}]{yamada:2010}
\bibinfo{author}{\bibfnamefont{M.}~\bibnamefont{Yamada}},
  \bibinfo{author}{\bibfnamefont{R.}~\bibnamefont{Kulsrud}}, \bibnamefont{and}
  \bibinfo{author}{\bibfnamefont{H.}~\bibnamefont{Ji}}, \bibinfo{journal}{Rev.
  Mod. Phys.} \textbf{\bibinfo{volume}{82}}, \bibinfo{pages}{603}
  (\bibinfo{year}{2010}).

\bibitem[{\citenamefont{{Mozer} and {Pritchett}}(2009)}]{mozer:2009grl}
\bibinfo{author}{\bibfnamefont{F.~S.} \bibnamefont{{Mozer}}} \bibnamefont{and}
  \bibinfo{author}{\bibfnamefont{P.~L.} \bibnamefont{{Pritchett}}},
  \bibinfo{journal}{\grl} \textbf{\bibinfo{volume}{36}}, \bibinfo{eid}{L07102}
  (\bibinfo{year}{2009}).

\bibitem[{\citenamefont{Graham et~al.}(2014)\citenamefont{Graham, Khotyaintsev,
  Vaivads, Andr{\'e}, and Fazakerley}}]{graham:2014}
\bibinfo{author}{\bibfnamefont{D.~B.} \bibnamefont{Graham}},
  \bibinfo{author}{\bibfnamefont{Y.~V.} \bibnamefont{Khotyaintsev}},
  \bibinfo{author}{\bibfnamefont{A.}~\bibnamefont{Vaivads}},
  \bibinfo{author}{\bibfnamefont{M.}~\bibnamefont{Andr{\'e}}},
  \bibnamefont{and}
  \bibinfo{author}{\bibfnamefont{A.}~\bibnamefont{Fazakerley}},
  \bibinfo{journal}{Physical Review Letters} \textbf{\bibinfo{volume}{112}},
  \bibinfo{pages}{215004} (\bibinfo{year}{2014}).

\bibitem[{\citenamefont{Hesse et~al.}(2014)\citenamefont{Hesse, Aunai, Sibeck,
  and Birn}}]{hesse:2014}
\bibinfo{author}{\bibfnamefont{M.}~\bibnamefont{Hesse}},
  \bibinfo{author}{\bibfnamefont{N.}~\bibnamefont{Aunai}},
  \bibinfo{author}{\bibfnamefont{D.}~\bibnamefont{Sibeck}}, \bibnamefont{and}
  \bibinfo{author}{\bibfnamefont{J.}~\bibnamefont{Birn}},
  \bibinfo{journal}{Geophysical Research Letters}
  \textbf{\bibinfo{volume}{41}}, \bibinfo{pages}{8673} (\bibinfo{year}{2014}).

\bibitem[{\citenamefont{Hesse et~al.}(2016)\citenamefont{Hesse, Liu, Chen,
  Bessho, Kuznetsova, Birn, and Burch}}]{hesse:2016}
\bibinfo{author}{\bibfnamefont{M.}~\bibnamefont{Hesse}},
  \bibinfo{author}{\bibfnamefont{Y.-H.} \bibnamefont{Liu}},
  \bibinfo{author}{\bibfnamefont{L.-J.} \bibnamefont{Chen}},
  \bibinfo{author}{\bibfnamefont{N.}~\bibnamefont{Bessho}},
  \bibinfo{author}{\bibfnamefont{M.}~\bibnamefont{Kuznetsova}},
  \bibinfo{author}{\bibfnamefont{J.}~\bibnamefont{Birn}}, \bibnamefont{and}
  \bibinfo{author}{\bibfnamefont{J.~L.} \bibnamefont{Burch}},
  \bibinfo{journal}{Geophysical Research Letters}
  \textbf{\bibinfo{volume}{43}}, \bibinfo{pages}{2359} (\bibinfo{year}{2016}).

\bibitem[{\citenamefont{Moore et~al.}(2013)\citenamefont{Moore, Burch,
  Daughton, Fuselier, Hasegawa, Petrinec, and Pu}}]{moore:2013}
\bibinfo{author}{\bibfnamefont{T.}~\bibnamefont{Moore}},
  \bibinfo{author}{\bibfnamefont{J.}~\bibnamefont{Burch}},
  \bibinfo{author}{\bibfnamefont{W.}~\bibnamefont{Daughton}},
  \bibinfo{author}{\bibfnamefont{S.}~\bibnamefont{Fuselier}},
  \bibinfo{author}{\bibfnamefont{H.}~\bibnamefont{Hasegawa}},
  \bibinfo{author}{\bibfnamefont{S.}~\bibnamefont{Petrinec}}, \bibnamefont{and}
  \bibinfo{author}{\bibfnamefont{Z.}~\bibnamefont{Pu}},
  \bibinfo{journal}{Journal of Atmospheric and Solar-Terrestrial Physics}
  \textbf{\bibinfo{volume}{99}}, \bibinfo{pages}{32} (\bibinfo{year}{2013}).

\bibitem[{\citenamefont{Sonnerup}(1974)}]{sonnerup:1974}
\bibinfo{author}{\bibfnamefont{B.~{\"O}.} \bibnamefont{Sonnerup}},
  \bibinfo{journal}{Journal of Geophysical Research}
  \textbf{\bibinfo{volume}{79}}, \bibinfo{pages}{1546} (\bibinfo{year}{1974}).

\bibitem[{\citenamefont{Russell and Elphic}(1979)}]{russell:1979}
\bibinfo{author}{\bibfnamefont{C.~T.} \bibnamefont{Russell}} \bibnamefont{and}
  \bibinfo{author}{\bibfnamefont{R.~C.} \bibnamefont{Elphic}},
  \bibinfo{journal}{Geophys. Res. Lett.} \textbf{\bibinfo{volume}{6}},
  \bibinfo{pages}{33} (\bibinfo{year}{1979}), ISSN \bibinfo{issn}{0094-8276},
  \urlprefix\url{http://dx.doi.org/10.1029/GL006i001p00033}.

\bibitem[{\citenamefont{Cowley}(1982)}]{cowley:1982}
\bibinfo{author}{\bibfnamefont{S.}~\bibnamefont{Cowley}},
  \bibinfo{journal}{Reviews of Geophysics} \textbf{\bibinfo{volume}{20}},
  \bibinfo{pages}{531} (\bibinfo{year}{1982}).

\bibitem[{\citenamefont{Wygant et~al.}(1983)\citenamefont{Wygant, Torbert, and
  Mozer}}]{wygant:1983}
\bibinfo{author}{\bibfnamefont{J.~R.} \bibnamefont{Wygant}},
  \bibinfo{author}{\bibfnamefont{R.~B.} \bibnamefont{Torbert}},
  \bibnamefont{and} \bibinfo{author}{\bibfnamefont{F.}~\bibnamefont{Mozer}},
  \bibinfo{journal}{Journal of Geophysical Research: Space Physics}
  \textbf{\bibinfo{volume}{88}}, \bibinfo{pages}{5727} (\bibinfo{year}{1983}).

\bibitem[{\citenamefont{Phan et~al.}(2000)\citenamefont{Phan, Kistler, Klecker,
  Haerendel, Paschmann, Sonnerup, Baumjohann, Bavassano-Cattaneo, Carlson,
  Dilellis et~al.}}]{phan:2000}
\bibinfo{author}{\bibfnamefont{T.~D.} \bibnamefont{Phan}},
  \bibinfo{author}{\bibfnamefont{L.~M.} \bibnamefont{Kistler}},
  \bibinfo{author}{\bibfnamefont{B.}~\bibnamefont{Klecker}},
  \bibinfo{author}{\bibfnamefont{G.}~\bibnamefont{Haerendel}},
  \bibinfo{author}{\bibfnamefont{G.}~\bibnamefont{Paschmann}},
  \bibinfo{author}{\bibfnamefont{B.~U.~O.} \bibnamefont{Sonnerup}},
  \bibinfo{author}{\bibfnamefont{W.}~\bibnamefont{Baumjohann}},
  \bibinfo{author}{\bibfnamefont{M.~B.} \bibnamefont{Bavassano-Cattaneo}},
  \bibinfo{author}{\bibfnamefont{C.~W.} \bibnamefont{Carlson}},
  \bibinfo{author}{\bibfnamefont{A.~M.} \bibnamefont{Dilellis}},
  \bibnamefont{et~al.}, \bibinfo{journal}{Nature}
  \textbf{\bibinfo{volume}{404}}, \bibinfo{pages}{848} (\bibinfo{year}{2000}),
  ISSN \bibinfo{issn}{0028-0836}.

\bibitem[{\citenamefont{Mozer et~al.}(2002)\citenamefont{Mozer, Bale, and
  Phan}}]{mozer:2002}
\bibinfo{author}{\bibfnamefont{F.~S.} \bibnamefont{Mozer}},
  \bibinfo{author}{\bibfnamefont{S.~D.} \bibnamefont{Bale}}, \bibnamefont{and}
  \bibinfo{author}{\bibfnamefont{T.~D.} \bibnamefont{Phan}},
  \bibinfo{journal}{Phys. Rev. Lett.} \textbf{\bibinfo{volume}{89}},
  \bibinfo{pages}{015002} (\bibinfo{year}{2002}),
  \urlprefix\url{http://link.aps.org/doi/10.1103/PhysRevLett.89.015002}.

\bibitem[{\citenamefont{{Paschmann} et~al.}(1979)\citenamefont{{Paschmann},
  {Papamastorakis}, {Sckopke}, {Haerendel}, {Sonnerup}, {Bame}, {Asbridge},
  {Gosling}, {Russel}, and {Elphic}}}]{paschmann:1979}
\bibinfo{author}{\bibfnamefont{G.}~\bibnamefont{{Paschmann}}},
  \bibinfo{author}{\bibfnamefont{I.}~\bibnamefont{{Papamastorakis}}},
  \bibinfo{author}{\bibfnamefont{N.}~\bibnamefont{{Sckopke}}},
  \bibinfo{author}{\bibfnamefont{G.}~\bibnamefont{{Haerendel}}},
  \bibinfo{author}{\bibfnamefont{B.~U.~O.} \bibnamefont{{Sonnerup}}},
  \bibinfo{author}{\bibfnamefont{S.~J.} \bibnamefont{{Bame}}},
  \bibinfo{author}{\bibfnamefont{J.~R.} \bibnamefont{{Asbridge}}},
  \bibinfo{author}{\bibfnamefont{J.~T.} \bibnamefont{{Gosling}}},
  \bibinfo{author}{\bibfnamefont{C.~T.} \bibnamefont{{Russel}}},
  \bibnamefont{and} \bibinfo{author}{\bibfnamefont{R.~C.}
  \bibnamefont{{Elphic}}}, \bibinfo{journal}{Nature}
  \textbf{\bibinfo{volume}{282}}, \bibinfo{pages}{243} (\bibinfo{year}{1979}).

\bibitem[{\citenamefont{Cassak and Shay}(2007)}]{cassak:2007}
\bibinfo{author}{\bibfnamefont{P.~A.} \bibnamefont{Cassak}} \bibnamefont{and}
  \bibinfo{author}{\bibfnamefont{M.~A.} \bibnamefont{Shay}},
  \bibinfo{journal}{Phys. Plasmas} \textbf{\bibinfo{volume}{14}}
  (\bibinfo{year}{2007}), ISSN \bibinfo{issn}{1070-664X}.

\bibitem[{\citenamefont{Borovsky and Hesse}(2007)}]{borovsky:2007}
\bibinfo{author}{\bibfnamefont{J.~E.} \bibnamefont{Borovsky}} \bibnamefont{and}
  \bibinfo{author}{\bibfnamefont{M.}~\bibnamefont{Hesse}},
  \bibinfo{journal}{Phys. Plasmas} \textbf{\bibinfo{volume}{14}}
  (\bibinfo{year}{2007}), ISSN \bibinfo{issn}{1070-664X}.

\bibitem[{\citenamefont{{Mozer} et~al.}(2008)\citenamefont{{Mozer},
  {Pritchett}, {Bonnell}, {Sundkvist}, and {Chang}}}]{mozer:2008}
\bibinfo{author}{\bibfnamefont{F.~S.} \bibnamefont{{Mozer}}},
  \bibinfo{author}{\bibfnamefont{P.~L.} \bibnamefont{{Pritchett}}},
  \bibinfo{author}{\bibfnamefont{J.}~\bibnamefont{{Bonnell}}},
  \bibinfo{author}{\bibfnamefont{D.}~\bibnamefont{{Sundkvist}}},
  \bibnamefont{and} \bibinfo{author}{\bibfnamefont{M.~T.}
  \bibnamefont{{Chang}}}, \bibinfo{journal}{Journal of Geophysical Research
  (Space Physics)} \textbf{\bibinfo{volume}{113}}, \bibinfo{eid}{A00C03}
  (\bibinfo{year}{2008}).

\bibitem[{\citenamefont{Pritchett}(2008)}]{pritchett:2008}
\bibinfo{author}{\bibfnamefont{P.~L.} \bibnamefont{Pritchett}},
  \bibinfo{journal}{J. Geophys. Res.} \textbf{\bibinfo{volume}{113}}
  (\bibinfo{year}{2008}), ISSN \bibinfo{issn}{0148-0227}.

\bibitem[{\citenamefont{Pritchett and Mozer}(2009)}]{pritchett:2009}
\bibinfo{author}{\bibfnamefont{P.~L.} \bibnamefont{Pritchett}}
  \bibnamefont{and} \bibinfo{author}{\bibfnamefont{F.~S.} \bibnamefont{Mozer}},
  \bibinfo{journal}{J. Geophys. Res.} \textbf{\bibinfo{volume}{114}}
  (\bibinfo{year}{2009}), ISSN \bibinfo{issn}{0148-0227}.

\bibitem[{\citenamefont{Bessho et~al.}(2016)\citenamefont{Bessho, Chen, and
  Hesse}}]{bessho:2016}
\bibinfo{author}{\bibfnamefont{N.}~\bibnamefont{Bessho}},
  \bibinfo{author}{\bibfnamefont{L.-J.} \bibnamefont{Chen}}, \bibnamefont{and}
  \bibinfo{author}{\bibfnamefont{M.}~\bibnamefont{Hesse}},
  \bibinfo{journal}{Geophysical Research Letters}
  \textbf{\bibinfo{volume}{43}}, \bibinfo{pages}{1828} (\bibinfo{year}{2016}).

\bibitem[{\citenamefont{Chen et~al.}(2016)\citenamefont{Chen, Hesse, Wang,
  Bessho, and Daughton}}]{chen:2016}
\bibinfo{author}{\bibfnamefont{L.-J.} \bibnamefont{Chen}},
  \bibinfo{author}{\bibfnamefont{M.}~\bibnamefont{Hesse}},
  \bibinfo{author}{\bibfnamefont{S.}~\bibnamefont{Wang}},
  \bibinfo{author}{\bibfnamefont{N.}~\bibnamefont{Bessho}}, \bibnamefont{and}
  \bibinfo{author}{\bibfnamefont{W.}~\bibnamefont{Daughton}},
  \bibinfo{journal}{Geophysical Research Letters}
  \textbf{\bibinfo{volume}{43}}, \bibinfo{pages}{2405} (\bibinfo{year}{2016}).

\bibitem[{\citenamefont{Egedal et~al.}(2016)\citenamefont{Egedal, Le, Daughton,
  Wetherton, Cassak, Chen, Lavraud, Torbert, Dorelli, Gershman
  et~al.}}]{egedal:2016}
\bibinfo{author}{\bibfnamefont{J.}~\bibnamefont{Egedal}},
  \bibinfo{author}{\bibfnamefont{A.}~\bibnamefont{Le}},
  \bibinfo{author}{\bibfnamefont{W.}~\bibnamefont{Daughton}},
  \bibinfo{author}{\bibfnamefont{B.}~\bibnamefont{Wetherton}},
  \bibinfo{author}{\bibfnamefont{P.}~\bibnamefont{Cassak}},
  \bibinfo{author}{\bibfnamefont{L.-J.} \bibnamefont{Chen}},
  \bibinfo{author}{\bibfnamefont{B.}~\bibnamefont{Lavraud}},
  \bibinfo{author}{\bibfnamefont{R.}~\bibnamefont{Torbert}},
  \bibinfo{author}{\bibfnamefont{J.}~\bibnamefont{Dorelli}},
  \bibinfo{author}{\bibfnamefont{D.}~\bibnamefont{Gershman}},
  \bibnamefont{et~al.}, \bibinfo{journal}{Physical review letters}
  \textbf{\bibinfo{volume}{117}}, \bibinfo{pages}{185101}
  (\bibinfo{year}{2016}).

\bibitem[{\citenamefont{Shay et~al.}(2016)\citenamefont{Shay, Phan, Haggerty,
  Fujimoto, Drake, Malakit, Cassak, and Swisdak}}]{shay:2016}
\bibinfo{author}{\bibfnamefont{M.}~\bibnamefont{Shay}},
  \bibinfo{author}{\bibfnamefont{T.}~\bibnamefont{Phan}},
  \bibinfo{author}{\bibfnamefont{C.}~\bibnamefont{Haggerty}},
  \bibinfo{author}{\bibfnamefont{M.}~\bibnamefont{Fujimoto}},
  \bibinfo{author}{\bibfnamefont{J.}~\bibnamefont{Drake}},
  \bibinfo{author}{\bibfnamefont{K.}~\bibnamefont{Malakit}},
  \bibinfo{author}{\bibfnamefont{P.}~\bibnamefont{Cassak}}, \bibnamefont{and}
  \bibinfo{author}{\bibfnamefont{M.}~\bibnamefont{Swisdak}},
  \bibinfo{journal}{Geophysical Research Letters}
  \textbf{\bibinfo{volume}{43}}, \bibinfo{pages}{4145} (\bibinfo{year}{2016}).

\bibitem[{\citenamefont{Torbert et~al.}(2016)\citenamefont{Torbert, Burch,
  Giles, Gershman, Pollock, Dorelli, Avanov, Argall, Shuster, Strangeway
  et~al.}}]{torbert:2016}
\bibinfo{author}{\bibfnamefont{R.}~\bibnamefont{Torbert}},
  \bibinfo{author}{\bibfnamefont{J.}~\bibnamefont{Burch}},
  \bibinfo{author}{\bibfnamefont{B.}~\bibnamefont{Giles}},
  \bibinfo{author}{\bibfnamefont{D.}~\bibnamefont{Gershman}},
  \bibinfo{author}{\bibfnamefont{C.}~\bibnamefont{Pollock}},
  \bibinfo{author}{\bibfnamefont{J.}~\bibnamefont{Dorelli}},
  \bibinfo{author}{\bibfnamefont{L.}~\bibnamefont{Avanov}},
  \bibinfo{author}{\bibfnamefont{M.}~\bibnamefont{Argall}},
  \bibinfo{author}{\bibfnamefont{J.}~\bibnamefont{Shuster}},
  \bibinfo{author}{\bibfnamefont{R.}~\bibnamefont{Strangeway}},
  \bibnamefont{et~al.}, \bibinfo{journal}{Geophysical Research Letters}
  \textbf{\bibinfo{volume}{43}}, \bibinfo{pages}{5918} (\bibinfo{year}{2016}).

\bibitem[{\citenamefont{Burch et~al.}(2016)\citenamefont{Burch, Torbert, Phan,
  Chen, Moore, Ergun, Eastwood, Gershman, Cassak, Argall
  et~al.}}]{burch:2016science}
\bibinfo{author}{\bibfnamefont{J.}~\bibnamefont{Burch}},
  \bibinfo{author}{\bibfnamefont{R.}~\bibnamefont{Torbert}},
  \bibinfo{author}{\bibfnamefont{T.}~\bibnamefont{Phan}},
  \bibinfo{author}{\bibfnamefont{L.-J.} \bibnamefont{Chen}},
  \bibinfo{author}{\bibfnamefont{T.}~\bibnamefont{Moore}},
  \bibinfo{author}{\bibfnamefont{R.}~\bibnamefont{Ergun}},
  \bibinfo{author}{\bibfnamefont{J.}~\bibnamefont{Eastwood}},
  \bibinfo{author}{\bibfnamefont{D.}~\bibnamefont{Gershman}},
  \bibinfo{author}{\bibfnamefont{P.}~\bibnamefont{Cassak}},
  \bibinfo{author}{\bibfnamefont{M.}~\bibnamefont{Argall}},
  \bibnamefont{et~al.}, \bibinfo{journal}{Science}
  \textbf{\bibinfo{volume}{352}}, \bibinfo{pages}{aaf2939}
  (\bibinfo{year}{2016}).

\bibitem[{\citenamefont{Burch and Phan}(2016)}]{burch:2016grl}
\bibinfo{author}{\bibfnamefont{J.}~\bibnamefont{Burch}} \bibnamefont{and}
  \bibinfo{author}{\bibfnamefont{T.}~\bibnamefont{Phan}},
  \bibinfo{journal}{Geophysical Research Letters}
  \textbf{\bibinfo{volume}{43}}, \bibinfo{pages}{8327} (\bibinfo{year}{2016}).

\bibitem[{\citenamefont{Zenitani et~al.}(2017)\citenamefont{Zenitani, Hasegawa,
  and Nagai}}]{zenitani:2017}
\bibinfo{author}{\bibfnamefont{S.}~\bibnamefont{Zenitani}},
  \bibinfo{author}{\bibfnamefont{H.}~\bibnamefont{Hasegawa}}, \bibnamefont{and}
  \bibinfo{author}{\bibfnamefont{T.}~\bibnamefont{Nagai}},
  \bibinfo{journal}{Journal of Geophysical Research: Space Physics}
  \textbf{\bibinfo{volume}{122}}, \bibinfo{pages}{7396} (\bibinfo{year}{2017}).

\bibitem[{\citenamefont{Egedal et~al.}(2018)\citenamefont{Egedal, Le, Daughton,
  Wetherton, Cassak, Burch, Lavraud, Dorelli, Gershman, and
  Avanov}}]{egedal:2018}
\bibinfo{author}{\bibfnamefont{J.}~\bibnamefont{Egedal}},
  \bibinfo{author}{\bibfnamefont{A.}~\bibnamefont{Le}},
  \bibinfo{author}{\bibfnamefont{W.}~\bibnamefont{Daughton}},
  \bibinfo{author}{\bibfnamefont{B.}~\bibnamefont{Wetherton}},
  \bibinfo{author}{\bibfnamefont{P.}~\bibnamefont{Cassak}},
  \bibinfo{author}{\bibfnamefont{J.}~\bibnamefont{Burch}},
  \bibinfo{author}{\bibfnamefont{B.}~\bibnamefont{Lavraud}},
  \bibinfo{author}{\bibfnamefont{J.}~\bibnamefont{Dorelli}},
  \bibinfo{author}{\bibfnamefont{D.}~\bibnamefont{Gershman}}, \bibnamefont{and}
  \bibinfo{author}{\bibfnamefont{L.}~\bibnamefont{Avanov}},
  \bibinfo{journal}{Physical Review Letters} \textbf{\bibinfo{volume}{120}},
  \bibinfo{pages}{055101} (\bibinfo{year}{2018}).

\bibitem[{\citenamefont{Davidson and Gladd}(1975)}]{davidson:1975}
\bibinfo{author}{\bibfnamefont{R.}~\bibnamefont{Davidson}} \bibnamefont{and}
  \bibinfo{author}{\bibfnamefont{N.}~\bibnamefont{Gladd}},
  \bibinfo{journal}{The Physics of Fluids} \textbf{\bibinfo{volume}{18}},
  \bibinfo{pages}{1327} (\bibinfo{year}{1975}).

\bibitem[{\citenamefont{Daughton}(2003)}]{daughton:2003}
\bibinfo{author}{\bibfnamefont{W.}~\bibnamefont{Daughton}},
  \bibinfo{journal}{Phys. Plasmas} \textbf{\bibinfo{volume}{10}},
  \bibinfo{pages}{3103} (\bibinfo{year}{2003}), ISSN \bibinfo{issn}{1070-664X}.

\bibitem[{\citenamefont{Vaivads et~al.}(2004)\citenamefont{Vaivads, Andr{\'e},
  Buchert, Wahlund, Fazakerley, and Cornilleau-Wehrlin}}]{vaivads:2004}
\bibinfo{author}{\bibfnamefont{A.}~\bibnamefont{Vaivads}},
  \bibinfo{author}{\bibfnamefont{M.}~\bibnamefont{Andr{\'e}}},
  \bibinfo{author}{\bibfnamefont{S.}~\bibnamefont{Buchert}},
  \bibinfo{author}{\bibfnamefont{J.-E.} \bibnamefont{Wahlund}},
  \bibinfo{author}{\bibfnamefont{A.}~\bibnamefont{Fazakerley}},
  \bibnamefont{and}
  \bibinfo{author}{\bibfnamefont{N.}~\bibnamefont{Cornilleau-Wehrlin}},
  \bibinfo{journal}{Geophysical research letters} \textbf{\bibinfo{volume}{31}}
  (\bibinfo{year}{2004}).

\bibitem[{\citenamefont{Graham et~al.}(2017)\citenamefont{Graham, Khotyaintsev,
  Norgren, Vaivads, Andr{\'e}, Toledo-Redondo, Lindqvist, Marklund, Ergun,
  Paterson et~al.}}]{graham:2017}
\bibinfo{author}{\bibfnamefont{D.~B.} \bibnamefont{Graham}},
  \bibinfo{author}{\bibfnamefont{Y.~V.} \bibnamefont{Khotyaintsev}},
  \bibinfo{author}{\bibfnamefont{C.}~\bibnamefont{Norgren}},
  \bibinfo{author}{\bibfnamefont{A.}~\bibnamefont{Vaivads}},
  \bibinfo{author}{\bibfnamefont{M.}~\bibnamefont{Andr{\'e}}},
  \bibinfo{author}{\bibfnamefont{S.}~\bibnamefont{Toledo-Redondo}},
  \bibinfo{author}{\bibfnamefont{P.-A.} \bibnamefont{Lindqvist}},
  \bibinfo{author}{\bibfnamefont{G.}~\bibnamefont{Marklund}},
  \bibinfo{author}{\bibfnamefont{R.}~\bibnamefont{Ergun}},
  \bibinfo{author}{\bibfnamefont{W.}~\bibnamefont{Paterson}},
  \bibnamefont{et~al.}, \bibinfo{journal}{Journal of Geophysical Research:
  Space Physics} \textbf{\bibinfo{volume}{122}}, \bibinfo{pages}{517}
  (\bibinfo{year}{2017}).

\bibitem[{\citenamefont{Carter et~al.}(2001)\citenamefont{Carter, Ji,
  Trintchouk, Yamada, and Kulsrud}}]{carter:2001}
\bibinfo{author}{\bibfnamefont{T.}~\bibnamefont{Carter}},
  \bibinfo{author}{\bibfnamefont{H.}~\bibnamefont{Ji}},
  \bibinfo{author}{\bibfnamefont{F.}~\bibnamefont{Trintchouk}},
  \bibinfo{author}{\bibfnamefont{M.}~\bibnamefont{Yamada}}, \bibnamefont{and}
  \bibinfo{author}{\bibfnamefont{R.}~\bibnamefont{Kulsrud}},
  \bibinfo{journal}{Physical review letters} \textbf{\bibinfo{volume}{88}},
  \bibinfo{pages}{015001} (\bibinfo{year}{2001}).

\bibitem[{\citenamefont{Yoo et~al.}(2014)\citenamefont{Yoo, Yamada, Ji,
  Jara-Almonte, Myers, and Chen}}]{yoo:2014}
\bibinfo{author}{\bibfnamefont{J.}~\bibnamefont{Yoo}},
  \bibinfo{author}{\bibfnamefont{M.}~\bibnamefont{Yamada}},
  \bibinfo{author}{\bibfnamefont{H.}~\bibnamefont{Ji}},
  \bibinfo{author}{\bibfnamefont{J.}~\bibnamefont{Jara-Almonte}},
  \bibinfo{author}{\bibfnamefont{C.~E.} \bibnamefont{Myers}}, \bibnamefont{and}
  \bibinfo{author}{\bibfnamefont{L.-J.} \bibnamefont{Chen}},
  \bibinfo{journal}{Physical review letters} \textbf{\bibinfo{volume}{113}},
  \bibinfo{pages}{095002} (\bibinfo{year}{2014}).

\bibitem[{\citenamefont{Yoo et~al.}(2017)\citenamefont{Yoo, Na, Jara-Almonte,
  Yamada, Ji, Roytershteyn, Argall, Fox, and Chen}}]{yoo:2017}
\bibinfo{author}{\bibfnamefont{J.}~\bibnamefont{Yoo}},
  \bibinfo{author}{\bibfnamefont{B.}~\bibnamefont{Na}},
  \bibinfo{author}{\bibfnamefont{J.}~\bibnamefont{Jara-Almonte}},
  \bibinfo{author}{\bibfnamefont{M.}~\bibnamefont{Yamada}},
  \bibinfo{author}{\bibfnamefont{H.}~\bibnamefont{Ji}},
  \bibinfo{author}{\bibfnamefont{V.}~\bibnamefont{Roytershteyn}},
  \bibinfo{author}{\bibfnamefont{M.~R.} \bibnamefont{Argall}},
  \bibinfo{author}{\bibfnamefont{W.}~\bibnamefont{Fox}}, \bibnamefont{and}
  \bibinfo{author}{\bibfnamefont{L.-J.} \bibnamefont{Chen}},
  \bibinfo{journal}{Journal of Geophysical Research: Space Physics}
  \textbf{\bibinfo{volume}{122}}, \bibinfo{pages}{9264} (\bibinfo{year}{2017}).

\bibitem[{\citenamefont{Bale et~al.}(2002)\citenamefont{Bale, Mozer, and
  Phan}}]{bale:2002}
\bibinfo{author}{\bibfnamefont{S.}~\bibnamefont{Bale}},
  \bibinfo{author}{\bibfnamefont{F.}~\bibnamefont{Mozer}}, \bibnamefont{and}
  \bibinfo{author}{\bibfnamefont{T.}~\bibnamefont{Phan}},
  \bibinfo{journal}{Geophysical research letters} \textbf{\bibinfo{volume}{29}}
  (\bibinfo{year}{2002}).

\bibitem[{\citenamefont{Norgren et~al.}(2012)\citenamefont{Norgren, Vaivads,
  Khotyaintsev, and Andr{\'e}}}]{norgren:2012}
\bibinfo{author}{\bibfnamefont{C.}~\bibnamefont{Norgren}},
  \bibinfo{author}{\bibfnamefont{A.}~\bibnamefont{Vaivads}},
  \bibinfo{author}{\bibfnamefont{Y.~V.} \bibnamefont{Khotyaintsev}},
  \bibnamefont{and}
  \bibinfo{author}{\bibfnamefont{M.}~\bibnamefont{Andr{\'e}}},
  \bibinfo{journal}{Physical review letters} \textbf{\bibinfo{volume}{109}},
  \bibinfo{pages}{055001} (\bibinfo{year}{2012}).

\bibitem[{\citenamefont{Treumann et~al.}(1991)\citenamefont{Treumann, LaBelle,
  and Pottelette}}]{treumann:1991}
\bibinfo{author}{\bibfnamefont{R.}~\bibnamefont{Treumann}},
  \bibinfo{author}{\bibfnamefont{J.}~\bibnamefont{LaBelle}}, \bibnamefont{and}
  \bibinfo{author}{\bibfnamefont{R.}~\bibnamefont{Pottelette}},
  \bibinfo{journal}{Journal of Geophysical Research: Space Physics}
  \textbf{\bibinfo{volume}{96}}, \bibinfo{pages}{16009} (\bibinfo{year}{1991}).

\bibitem[{\citenamefont{Gary and Sgro}(1990)}]{gary:1990}
\bibinfo{author}{\bibfnamefont{S.~P.} \bibnamefont{Gary}} \bibnamefont{and}
  \bibinfo{author}{\bibfnamefont{A.}~\bibnamefont{Sgro}},
  \bibinfo{journal}{Geophysical Research Letters}
  \textbf{\bibinfo{volume}{17}}, \bibinfo{pages}{909} (\bibinfo{year}{1990}).

\bibitem[{\citenamefont{Le et~al.}(2017)\citenamefont{Le, Daughton, Chen, and
  Egedal}}]{le:2017}
\bibinfo{author}{\bibfnamefont{A.}~\bibnamefont{Le}},
  \bibinfo{author}{\bibfnamefont{W.}~\bibnamefont{Daughton}},
  \bibinfo{author}{\bibfnamefont{L.-J.} \bibnamefont{Chen}}, \bibnamefont{and}
  \bibinfo{author}{\bibfnamefont{J.}~\bibnamefont{Egedal}},
  \bibinfo{journal}{Geophysical Research Letters}
  \textbf{\bibinfo{volume}{44}}, \bibinfo{pages}{2096} (\bibinfo{year}{2017}).

\bibitem[{\citenamefont{{Huba} et~al.}(1977)\citenamefont{{Huba}, {Gladd}, and
  {Papadopoulos}}}]{huba:1977}
\bibinfo{author}{\bibfnamefont{J.~D.} \bibnamefont{{Huba}}},
  \bibinfo{author}{\bibfnamefont{N.~T.} \bibnamefont{{Gladd}}},
  \bibnamefont{and}
  \bibinfo{author}{\bibfnamefont{K.}~\bibnamefont{{Papadopoulos}}},
  \bibinfo{journal}{\grl} \textbf{\bibinfo{volume}{4}}, \bibinfo{pages}{125}
  (\bibinfo{year}{1977}).

\bibitem[{\citenamefont{Davidson et~al.}(1977)\citenamefont{Davidson, Gladd,
  Wu, and Huba}}]{davidson:1977}
\bibinfo{author}{\bibfnamefont{R.}~\bibnamefont{Davidson}},
  \bibinfo{author}{\bibfnamefont{N.}~\bibnamefont{Gladd}},
  \bibinfo{author}{\bibfnamefont{C.}~\bibnamefont{Wu}}, \bibnamefont{and}
  \bibinfo{author}{\bibfnamefont{J.}~\bibnamefont{Huba}}, \bibinfo{journal}{The
  Physics of Fluids} \textbf{\bibinfo{volume}{20}}, \bibinfo{pages}{301}
  (\bibinfo{year}{1977}).

\bibitem[{\citenamefont{Roytershteyn et~al.}(2012)\citenamefont{Roytershteyn,
  Daughton, Karimabadi, and Mozer}}]{roytershteyn:2012}
\bibinfo{author}{\bibfnamefont{V.}~\bibnamefont{Roytershteyn}},
  \bibinfo{author}{\bibfnamefont{W.}~\bibnamefont{Daughton}},
  \bibinfo{author}{\bibfnamefont{H.}~\bibnamefont{Karimabadi}},
  \bibnamefont{and} \bibinfo{author}{\bibfnamefont{F.~S.} \bibnamefont{Mozer}},
  \bibinfo{journal}{Phys. Rev. Lett.} \textbf{\bibinfo{volume}{108}},
  \bibinfo{pages}{185001} (\bibinfo{year}{2012}),
  \urlprefix\url{http://link.aps.org/doi/10.1103/PhysRevLett.108.185001}.

\bibitem[{\citenamefont{Price et~al.}(2016)\citenamefont{Price, Swisdak, Drake,
  Cassak, Dahlin, and Ergun}}]{price:2016}
\bibinfo{author}{\bibfnamefont{L.}~\bibnamefont{Price}},
  \bibinfo{author}{\bibfnamefont{M.}~\bibnamefont{Swisdak}},
  \bibinfo{author}{\bibfnamefont{J.}~\bibnamefont{Drake}},
  \bibinfo{author}{\bibfnamefont{P.}~\bibnamefont{Cassak}},
  \bibinfo{author}{\bibfnamefont{J.}~\bibnamefont{Dahlin}}, \bibnamefont{and}
  \bibinfo{author}{\bibfnamefont{R.}~\bibnamefont{Ergun}},
  \bibinfo{journal}{Geophysical Research Letters}
  \textbf{\bibinfo{volume}{43}}, \bibinfo{pages}{6020} (\bibinfo{year}{2016}).

\bibitem[{\citenamefont{Price et~al.}(2017)\citenamefont{Price, Swisdak, Drake,
  Burch, Cassak, and Ergun}}]{price:2017}
\bibinfo{author}{\bibfnamefont{L.}~\bibnamefont{Price}},
  \bibinfo{author}{\bibfnamefont{M.}~\bibnamefont{Swisdak}},
  \bibinfo{author}{\bibfnamefont{J.}~\bibnamefont{Drake}},
  \bibinfo{author}{\bibfnamefont{J.}~\bibnamefont{Burch}},
  \bibinfo{author}{\bibfnamefont{P.}~\bibnamefont{Cassak}}, \bibnamefont{and}
  \bibinfo{author}{\bibfnamefont{R.}~\bibnamefont{Ergun}},
  \bibinfo{journal}{Journal of Geophysical Research: Space Physics}
  (\bibinfo{year}{2017}).

\bibitem[{\citenamefont{Che et~al.}(2011)\citenamefont{Che, Drake, and
  Swisdak}}]{che:2011}
\bibinfo{author}{\bibfnamefont{H.}~\bibnamefont{Che}},
  \bibinfo{author}{\bibfnamefont{J.}~\bibnamefont{Drake}}, \bibnamefont{and}
  \bibinfo{author}{\bibfnamefont{M.}~\bibnamefont{Swisdak}},
  \bibinfo{journal}{Nature} \textbf{\bibinfo{volume}{474}},
  \bibinfo{pages}{184} (\bibinfo{year}{2011}).

\bibitem[{\citenamefont{Chen et~al.}(2017)\citenamefont{Chen, Hesse, Wang,
  Gershman, Ergun, Burch, Bessho, Torbert, Giles, Webster et~al.}}]{chen:2017}
\bibinfo{author}{\bibfnamefont{L.-J.} \bibnamefont{Chen}},
  \bibinfo{author}{\bibfnamefont{M.}~\bibnamefont{Hesse}},
  \bibinfo{author}{\bibfnamefont{S.}~\bibnamefont{Wang}},
  \bibinfo{author}{\bibfnamefont{D.}~\bibnamefont{Gershman}},
  \bibinfo{author}{\bibfnamefont{R.}~\bibnamefont{Ergun}},
  \bibinfo{author}{\bibfnamefont{J.}~\bibnamefont{Burch}},
  \bibinfo{author}{\bibfnamefont{N.}~\bibnamefont{Bessho}},
  \bibinfo{author}{\bibfnamefont{R.}~\bibnamefont{Torbert}},
  \bibinfo{author}{\bibfnamefont{B.}~\bibnamefont{Giles}},
  \bibinfo{author}{\bibfnamefont{J.}~\bibnamefont{Webster}},
  \bibnamefont{et~al.}, \bibinfo{journal}{Journal of Geophysical Research:
  Space Physics}  (\bibinfo{year}{2017}).

\bibitem[{\citenamefont{Bowers et~al.}(2008)\citenamefont{Bowers, Albright,
  Yin, Bergen, and Kwan}}]{bowers:2008}
\bibinfo{author}{\bibfnamefont{K.~J.} \bibnamefont{Bowers}},
  \bibinfo{author}{\bibfnamefont{B.~J.} \bibnamefont{Albright}},
  \bibinfo{author}{\bibfnamefont{L.}~\bibnamefont{Yin}},
  \bibinfo{author}{\bibfnamefont{B.}~\bibnamefont{Bergen}}, \bibnamefont{and}
  \bibinfo{author}{\bibfnamefont{T.~J.~T.} \bibnamefont{Kwan}},
  \bibinfo{journal}{Physics of Plasmas} \textbf{\bibinfo{volume}{15}},
  \bibinfo{eid}{055703} (pages~\bibinfo{numpages}{7}) (\bibinfo{year}{2008}),
  \urlprefix\url{http://link.aip.org/link/?PHP/15/055703/1}.

\bibitem[{\citenamefont{Nakamura et~al.}(2017)\citenamefont{Nakamura, Eriksson,
  Hasegawa, Zenitani, Li, Genestreti, Nakamura, and Daughton}}]{nakamura:2017}
\bibinfo{author}{\bibfnamefont{T.}~\bibnamefont{Nakamura}},
  \bibinfo{author}{\bibfnamefont{S.}~\bibnamefont{Eriksson}},
  \bibinfo{author}{\bibfnamefont{H.}~\bibnamefont{Hasegawa}},
  \bibinfo{author}{\bibfnamefont{S.}~\bibnamefont{Zenitani}},
  \bibinfo{author}{\bibfnamefont{W.}~\bibnamefont{Li}},
  \bibinfo{author}{\bibfnamefont{K.}~\bibnamefont{Genestreti}},
  \bibinfo{author}{\bibfnamefont{R.}~\bibnamefont{Nakamura}}, \bibnamefont{and}
  \bibinfo{author}{\bibfnamefont{W.}~\bibnamefont{Daughton}},
  \bibinfo{journal}{Journal of Geophysical Research: Space Physics}
  \textbf{\bibinfo{volume}{122}} (\bibinfo{year}{2017}).

\bibitem[{\citenamefont{Daughton et~al.}(2014)\citenamefont{Daughton, Nakamura,
  Karimabadi, Roytershteyn, and Loring}}]{daughton:2014}
\bibinfo{author}{\bibfnamefont{W.}~\bibnamefont{Daughton}},
  \bibinfo{author}{\bibfnamefont{T.}~\bibnamefont{Nakamura}},
  \bibinfo{author}{\bibfnamefont{H.}~\bibnamefont{Karimabadi}},
  \bibinfo{author}{\bibfnamefont{V.}~\bibnamefont{Roytershteyn}},
  \bibnamefont{and} \bibinfo{author}{\bibfnamefont{B.}~\bibnamefont{Loring}},
  \bibinfo{journal}{Physics of Plasmas} \textbf{\bibinfo{volume}{21}},
  \bibinfo{pages}{052307} (\bibinfo{year}{2014}).

\bibitem[{\citenamefont{Che}(2017)}]{che:2017}
\bibinfo{author}{\bibfnamefont{H.}~\bibnamefont{Che}},
  \bibinfo{journal}{Physics of Plasmas} \textbf{\bibinfo{volume}{24}},
  \bibinfo{pages}{082115} (\bibinfo{year}{2017}),
  \eprint{http://dx.doi.org/10.1063/1.5000071},
  \urlprefix\url{http://dx.doi.org/10.1063/1.5000071}.

\bibitem[{\citenamefont{Winske et~al.}(1995)\citenamefont{Winske, Thomas, and
  Omidi}}]{winske:1995}
\bibinfo{author}{\bibfnamefont{D.}~\bibnamefont{Winske}},
  \bibinfo{author}{\bibfnamefont{V.}~\bibnamefont{Thomas}}, \bibnamefont{and}
  \bibinfo{author}{\bibfnamefont{N.}~\bibnamefont{Omidi}},
  \bibinfo{journal}{Physics of the Magnetopause} pp. \bibinfo{pages}{321--330}
  (\bibinfo{year}{1995}).

\bibitem[{\citenamefont{Winske and Liewer}(1978)}]{winske:1978}
\bibinfo{author}{\bibfnamefont{D.}~\bibnamefont{Winske}} \bibnamefont{and}
  \bibinfo{author}{\bibfnamefont{P.}~\bibnamefont{Liewer}},
  \bibinfo{journal}{The Physics of Fluids} \textbf{\bibinfo{volume}{21}},
  \bibinfo{pages}{1017} (\bibinfo{year}{1978}).

\bibitem[{\citenamefont{{Priest} and {D{\'e}moulin}}(1995)}]{priest:1995}
\bibinfo{author}{\bibfnamefont{E.~R.} \bibnamefont{{Priest}}} \bibnamefont{and}
  \bibinfo{author}{\bibfnamefont{P.}~\bibnamefont{{D{\'e}moulin}}},
  \bibinfo{journal}{\jgr} \textbf{\bibinfo{volume}{100}},
  \bibinfo{pages}{23443} (\bibinfo{year}{1995}).

\bibitem[{\citenamefont{Titov et~al.}(2002)\citenamefont{Titov, Hornig, and
  D{\'e}moulin}}]{titov:2002}
\bibinfo{author}{\bibfnamefont{V.~S.} \bibnamefont{Titov}},
  \bibinfo{author}{\bibfnamefont{G.}~\bibnamefont{Hornig}}, \bibnamefont{and}
  \bibinfo{author}{\bibfnamefont{P.}~\bibnamefont{D{\'e}moulin}},
  \bibinfo{journal}{Journal of Geophysical Research: Space Physics}
  \textbf{\bibinfo{volume}{107}} (\bibinfo{year}{2002}).

\bibitem[{\citenamefont{Borgogno et~al.}(2011)\citenamefont{Borgogno, Grasso,
  Pegoraro, and Schep}}]{borgogno:2011}
\bibinfo{author}{\bibfnamefont{D.}~\bibnamefont{Borgogno}},
  \bibinfo{author}{\bibfnamefont{D.}~\bibnamefont{Grasso}},
  \bibinfo{author}{\bibfnamefont{F.}~\bibnamefont{Pegoraro}}, \bibnamefont{and}
  \bibinfo{author}{\bibfnamefont{T.}~\bibnamefont{Schep}},
  \bibinfo{journal}{Physics of Plasmas} \textbf{\bibinfo{volume}{18}},
  \bibinfo{pages}{102307} (\bibinfo{year}{2011}).

\bibitem[{\citenamefont{Liu et~al.}(2017)\citenamefont{Liu, Hesse, Cassak,
  Shay, Wang, and Chen}}]{liu:2017}
\bibinfo{author}{\bibfnamefont{Y.-H.} \bibnamefont{Liu}},
  \bibinfo{author}{\bibfnamefont{M.}~\bibnamefont{Hesse}},
  \bibinfo{author}{\bibfnamefont{P.}~\bibnamefont{Cassak}},
  \bibinfo{author}{\bibfnamefont{M.}~\bibnamefont{Shay}},
  \bibinfo{author}{\bibfnamefont{S.}~\bibnamefont{Wang}}, \bibnamefont{and}
  \bibinfo{author}{\bibfnamefont{L.-J.} \bibnamefont{Chen}},
  \bibinfo{journal}{arXiv preprint arXiv:1711.06708}  (\bibinfo{year}{2017}).

\bibitem[{\citenamefont{Le et~al.}({2009})\citenamefont{Le, Egedal, Daughton,
  Fox, and Katz}}]{le:2009}
\bibinfo{author}{\bibfnamefont{A.}~\bibnamefont{Le}},
  \bibinfo{author}{\bibfnamefont{J.}~\bibnamefont{Egedal}},
  \bibinfo{author}{\bibfnamefont{W.}~\bibnamefont{Daughton}},
  \bibinfo{author}{\bibfnamefont{W.}~\bibnamefont{Fox}}, \bibnamefont{and}
  \bibinfo{author}{\bibfnamefont{N.}~\bibnamefont{Katz}},
  \bibinfo{journal}{Phys. Rev. Lett.} \textbf{\bibinfo{volume}{{102}}},
  \bibinfo{pages}{{085001}} (\bibinfo{year}{{2009}}), ISSN
  \bibinfo{issn}{{0031-9007}}.

\bibitem[{\citenamefont{Egedal et~al.}(2011)\citenamefont{Egedal, Le,
  Pritchett, and Daughton}}]{egedal:2011pop}
\bibinfo{author}{\bibfnamefont{J.}~\bibnamefont{Egedal}},
  \bibinfo{author}{\bibfnamefont{A.}~\bibnamefont{Le}},
  \bibinfo{author}{\bibfnamefont{P.~L.} \bibnamefont{Pritchett}},
  \bibnamefont{and} \bibinfo{author}{\bibfnamefont{W.}~\bibnamefont{Daughton}},
  \bibinfo{journal}{Physics of Plasmas} \textbf{\bibinfo{volume}{18}},
  \bibinfo{eid}{102901} (pages~\bibinfo{numpages}{8}) (\bibinfo{year}{2011}),
  \urlprefix\url{http://link.aip.org/link/?PHP/18/102901/1}.

\bibitem[{\citenamefont{Egedal et~al.}({2013})\citenamefont{Egedal, Le, and
  Daughton}}]{egedal:2013pop}
\bibinfo{author}{\bibfnamefont{J.}~\bibnamefont{Egedal}},
  \bibinfo{author}{\bibfnamefont{A.}~\bibnamefont{Le}}, \bibnamefont{and}
  \bibinfo{author}{\bibfnamefont{W.}~\bibnamefont{Daughton}},
  \bibinfo{journal}{Phys. Plasmas} \textbf{\bibinfo{volume}{{20}}}
  (\bibinfo{year}{{2013}}), ISSN \bibinfo{issn}{{1070-664X}}.

\bibitem[{\citenamefont{Chew et~al.}(1956)\citenamefont{Chew, Goldberger, and
  Low}}]{chew:1956}
\bibinfo{author}{\bibfnamefont{G.~F.} \bibnamefont{Chew}},
  \bibinfo{author}{\bibfnamefont{M.~L.} \bibnamefont{Goldberger}},
  \bibnamefont{and} \bibinfo{author}{\bibfnamefont{F.~E.} \bibnamefont{Low}},
  \bibinfo{journal}{Proc.~Royal Soc.~A} \textbf{\bibinfo{volume}{112}},
  \bibinfo{pages}{236} (\bibinfo{year}{1956}).

\bibitem[{\citenamefont{Wang et~al.}(2017)\citenamefont{Wang, Chen, Hesse,
  Wilson, Bessho, Gershman, Ergun, Phan, Burch, Dorelli et~al.}}]{wang:2017}
\bibinfo{author}{\bibfnamefont{S.}~\bibnamefont{Wang}},
  \bibinfo{author}{\bibfnamefont{L.-J.} \bibnamefont{Chen}},
  \bibinfo{author}{\bibfnamefont{M.}~\bibnamefont{Hesse}},
  \bibinfo{author}{\bibfnamefont{L.~B.} \bibnamefont{Wilson}},
  \bibinfo{author}{\bibfnamefont{N.}~\bibnamefont{Bessho}},
  \bibinfo{author}{\bibfnamefont{D.~J.} \bibnamefont{Gershman}},
  \bibinfo{author}{\bibfnamefont{R.~E.} \bibnamefont{Ergun}},
  \bibinfo{author}{\bibfnamefont{T.~D.} \bibnamefont{Phan}},
  \bibinfo{author}{\bibfnamefont{J.~L.} \bibnamefont{Burch}},
  \bibinfo{author}{\bibfnamefont{J.~C.} \bibnamefont{Dorelli}},
  \bibnamefont{et~al.}, \bibinfo{journal}{Geophysical Research Letters}
  (\bibinfo{year}{2017}).

\bibitem[{\citenamefont{Parker}(1957)}]{parker:1957}
\bibinfo{author}{\bibfnamefont{E.~N.} \bibnamefont{Parker}},
  \bibinfo{journal}{J. Geophys. Res.} \textbf{\bibinfo{volume}{62}},
  \bibinfo{pages}{509} (\bibinfo{year}{1957}).

\bibitem[{\citenamefont{Yoon and Lui}(2007)}]{yoon:2007}
\bibinfo{author}{\bibfnamefont{P.~H.} \bibnamefont{Yoon}} \bibnamefont{and}
  \bibinfo{author}{\bibfnamefont{A.~T.} \bibnamefont{Lui}},
  \bibinfo{journal}{Journal of Geophysical Research: Space Physics}
  \textbf{\bibinfo{volume}{112}} (\bibinfo{year}{2007}).

\bibitem[{\citenamefont{Kleva et~al.}(1995)\citenamefont{Kleva, Drake, and
  Waelbroeck}}]{kleva:1995}
\bibinfo{author}{\bibfnamefont{R.~G.} \bibnamefont{Kleva}},
  \bibinfo{author}{\bibfnamefont{J.~F.} \bibnamefont{Drake}}, \bibnamefont{and}
  \bibinfo{author}{\bibfnamefont{F.~L.} \bibnamefont{Waelbroeck}},
  \bibinfo{journal}{Phys. Plasmas} \textbf{\bibinfo{volume}{2}},
  \bibinfo{pages}{23} (\bibinfo{year}{1995}), ISSN \bibinfo{issn}{1070-664X}.

\bibitem[{\citenamefont{Ma and Bhattacharjee}(1996)}]{ma:1996}
\bibinfo{author}{\bibfnamefont{Z.}~\bibnamefont{Ma}} \bibnamefont{and}
  \bibinfo{author}{\bibfnamefont{A.}~\bibnamefont{Bhattacharjee}},
  \bibinfo{journal}{Geophysical research letters}
  \textbf{\bibinfo{volume}{23}}, \bibinfo{pages}{1673} (\bibinfo{year}{1996}).

\bibitem[{\citenamefont{Biskamp et~al.}(1997)\citenamefont{Biskamp, Schwarz,
  and Drake}}]{biskamp:1997}
\bibinfo{author}{\bibfnamefont{D.}~\bibnamefont{Biskamp}},
  \bibinfo{author}{\bibfnamefont{E.}~\bibnamefont{Schwarz}}, \bibnamefont{and}
  \bibinfo{author}{\bibfnamefont{J.~F.} \bibnamefont{Drake}},
  \bibinfo{journal}{Phys. Plasmas} \textbf{\bibinfo{volume}{4}},
  \bibinfo{pages}{1002} (\bibinfo{year}{1997}), ISSN \bibinfo{issn}{1070-664X}.

\bibitem[{\citenamefont{Kuznetsova et~al.}(2001)\citenamefont{Kuznetsova,
  Hesse, and Winske}}]{kuznetsova:2001}
\bibinfo{author}{\bibfnamefont{M.~M.} \bibnamefont{Kuznetsova}},
  \bibinfo{author}{\bibfnamefont{M.}~\bibnamefont{Hesse}}, \bibnamefont{and}
  \bibinfo{author}{\bibfnamefont{D.}~\bibnamefont{Winske}},
  \bibinfo{journal}{Journal of Geophysical Research: Space Physics
  (1978--2012)} \textbf{\bibinfo{volume}{106}}, \bibinfo{pages}{3799}
  (\bibinfo{year}{2001}).

\bibitem[{\citenamefont{Birn et~al.}(2001)\citenamefont{Birn, Drake, Shay,
  Rogers, Denton, Hesse, Kuznetsova, Ma, Bhattacharjee, Otto
  et~al.}}]{birn:2001}
\bibinfo{author}{\bibfnamefont{J.}~\bibnamefont{Birn}},
  \bibinfo{author}{\bibfnamefont{J.~F.} \bibnamefont{Drake}},
  \bibinfo{author}{\bibfnamefont{M.~A.} \bibnamefont{Shay}},
  \bibinfo{author}{\bibfnamefont{B.~N.} \bibnamefont{Rogers}},
  \bibinfo{author}{\bibfnamefont{R.~E.} \bibnamefont{Denton}},
  \bibinfo{author}{\bibfnamefont{M.}~\bibnamefont{Hesse}},
  \bibinfo{author}{\bibfnamefont{M.}~\bibnamefont{Kuznetsova}},
  \bibinfo{author}{\bibfnamefont{Z.~W.} \bibnamefont{Ma}},
  \bibinfo{author}{\bibfnamefont{A.}~\bibnamefont{Bhattacharjee}},
  \bibinfo{author}{\bibfnamefont{A.}~\bibnamefont{Otto}}, \bibnamefont{et~al.},
  \bibinfo{journal}{J. Geophys. Res.} \textbf{\bibinfo{volume}{106}},
  \bibinfo{pages}{3715} (\bibinfo{year}{2001}), ISSN \bibinfo{issn}{0148-0227}.

\bibitem[{\citenamefont{Daughton et~al.}(2006)\citenamefont{Daughton, Scudder,
  and Karimabadi}}]{daughton:2006}
\bibinfo{author}{\bibfnamefont{W.}~\bibnamefont{Daughton}},
  \bibinfo{author}{\bibfnamefont{J.}~\bibnamefont{Scudder}}, \bibnamefont{and}
  \bibinfo{author}{\bibfnamefont{H.}~\bibnamefont{Karimabadi}},
  \bibinfo{journal}{Phys. Plasmas} \textbf{\bibinfo{volume}{13}},
  \bibinfo{pages}{072101} (\bibinfo{year}{2006}), ISSN
  \bibinfo{issn}{1070-664X}.

\bibitem[{\citenamefont{Cassak et~al.}(2007)\citenamefont{Cassak, Drake, Shay,
  and Eckhardt}}]{cassak:2007prl}
\bibinfo{author}{\bibfnamefont{P.}~\bibnamefont{Cassak}},
  \bibinfo{author}{\bibfnamefont{J.}~\bibnamefont{Drake}},
  \bibinfo{author}{\bibfnamefont{M.}~\bibnamefont{Shay}}, \bibnamefont{and}
  \bibinfo{author}{\bibfnamefont{B.}~\bibnamefont{Eckhardt}},
  \bibinfo{journal}{Physical review letters} \textbf{\bibinfo{volume}{98}},
  \bibinfo{pages}{215001} (\bibinfo{year}{2007}).

\bibitem[{\citenamefont{Ozaki et~al.}(1996)\citenamefont{Ozaki, Sato, Horiuchi,
  and Group}}]{ozaki:1996}
\bibinfo{author}{\bibfnamefont{M.}~\bibnamefont{Ozaki}},
  \bibinfo{author}{\bibfnamefont{T.}~\bibnamefont{Sato}},
  \bibinfo{author}{\bibfnamefont{R.}~\bibnamefont{Horiuchi}}, \bibnamefont{and}
  \bibinfo{author}{\bibfnamefont{C.~S.} \bibnamefont{Group}},
  \bibinfo{journal}{Physics of Plasmas} \textbf{\bibinfo{volume}{3}},
  \bibinfo{pages}{2265} (\bibinfo{year}{1996}).

\bibitem[{\citenamefont{Tanaka and Sato}(1981)}]{tanaka:1981}
\bibinfo{author}{\bibfnamefont{M.}~\bibnamefont{Tanaka}} \bibnamefont{and}
  \bibinfo{author}{\bibfnamefont{T.}~\bibnamefont{Sato}},
  \bibinfo{journal}{Journal of Geophysical Research: Space Physics}
  \textbf{\bibinfo{volume}{86}}, \bibinfo{pages}{5541} (\bibinfo{year}{1981}).

\bibitem[{\citenamefont{Kulsrud et~al.}(2005)\citenamefont{Kulsrud, Ji, Fox,
  and Yamada}}]{kulsrud:2005}
\bibinfo{author}{\bibfnamefont{R.}~\bibnamefont{Kulsrud}},
  \bibinfo{author}{\bibfnamefont{H.}~\bibnamefont{Ji}},
  \bibinfo{author}{\bibfnamefont{W.}~\bibnamefont{Fox}}, \bibnamefont{and}
  \bibinfo{author}{\bibfnamefont{M.}~\bibnamefont{Yamada}},
  \bibinfo{journal}{Phys. Plasmas} \textbf{\bibinfo{volume}{12}},
  \bibinfo{pages}{082301} (\bibinfo{year}{2005}), ISSN
  \bibinfo{issn}{1070-664X}.

\bibitem[{\citenamefont{Krall}(1977)}]{krall:1977}
\bibinfo{author}{\bibfnamefont{N.}~\bibnamefont{Krall}}, \bibinfo{journal}{The
  Physics of Fluids} \textbf{\bibinfo{volume}{20}}, \bibinfo{pages}{311}
  (\bibinfo{year}{1977}).

\bibitem[{\citenamefont{Brackbill et~al.}(1984)\citenamefont{Brackbill,
  Forslund, Quest, and Winske}}]{brackbill:1984}
\bibinfo{author}{\bibfnamefont{J.}~\bibnamefont{Brackbill}},
  \bibinfo{author}{\bibfnamefont{D.}~\bibnamefont{Forslund}},
  \bibinfo{author}{\bibfnamefont{K.}~\bibnamefont{Quest}}, \bibnamefont{and}
  \bibinfo{author}{\bibfnamefont{D.}~\bibnamefont{Winske}},
  \bibinfo{journal}{The Physics of fluids} \textbf{\bibinfo{volume}{27}},
  \bibinfo{pages}{2682} (\bibinfo{year}{1984}).

\bibitem[{\citenamefont{Pritchett et~al.}(1996)\citenamefont{Pritchett,
  Coroniti, and Decyk}}]{pritchett:1996}
\bibinfo{author}{\bibfnamefont{P.}~\bibnamefont{Pritchett}},
  \bibinfo{author}{\bibfnamefont{F.}~\bibnamefont{Coroniti}}, \bibnamefont{and}
  \bibinfo{author}{\bibfnamefont{V.}~\bibnamefont{Decyk}},
  \bibinfo{journal}{Journal of Geophysical Research: Space Physics}
  \textbf{\bibinfo{volume}{101}}, \bibinfo{pages}{27413}
  (\bibinfo{year}{1996}).

\bibitem[{\citenamefont{Lapenta and Brackbill}(2002)}]{lapenta:2002}
\bibinfo{author}{\bibfnamefont{G.}~\bibnamefont{Lapenta}} \bibnamefont{and}
  \bibinfo{author}{\bibfnamefont{J.}~\bibnamefont{Brackbill}},
  \bibinfo{journal}{Physics of Plasmas} \textbf{\bibinfo{volume}{9}},
  \bibinfo{pages}{1544} (\bibinfo{year}{2002}).

\bibitem[{\citenamefont{Daughton et~al.}(2004)\citenamefont{Daughton, Lapenta,
  and Ricci}}]{daughton:2004}
\bibinfo{author}{\bibfnamefont{W.}~\bibnamefont{Daughton}},
  \bibinfo{author}{\bibfnamefont{G.}~\bibnamefont{Lapenta}}, \bibnamefont{and}
  \bibinfo{author}{\bibfnamefont{P.}~\bibnamefont{Ricci}},
  \bibinfo{journal}{Physical review letters} \textbf{\bibinfo{volume}{93}},
  \bibinfo{pages}{105004} (\bibinfo{year}{2004}).

\bibitem[{\citenamefont{Silin et~al.}(2005)\citenamefont{Silin, B{\"u}chner,
  and Vaivads}}]{silin:2005}
\bibinfo{author}{\bibfnamefont{I.}~\bibnamefont{Silin}},
  \bibinfo{author}{\bibfnamefont{J.}~\bibnamefont{B{\"u}chner}},
  \bibnamefont{and} \bibinfo{author}{\bibfnamefont{A.}~\bibnamefont{Vaivads}},
  \bibinfo{journal}{Physics of plasmas} \textbf{\bibinfo{volume}{12}},
  \bibinfo{pages}{062902} (\bibinfo{year}{2005}).

\bibitem[{\citenamefont{Ji et~al.}(2004)\citenamefont{Ji, Terry, Yamada,
  Kulsrud, Kuritsyn, and Ren}}]{ji:2004}
\bibinfo{author}{\bibfnamefont{H.~T.} \bibnamefont{Ji}},
  \bibinfo{author}{\bibfnamefont{S.}~\bibnamefont{Terry}},
  \bibinfo{author}{\bibfnamefont{M.}~\bibnamefont{Yamada}},
  \bibinfo{author}{\bibfnamefont{R.}~\bibnamefont{Kulsrud}},
  \bibinfo{author}{\bibfnamefont{A.}~\bibnamefont{Kuritsyn}}, \bibnamefont{and}
  \bibinfo{author}{\bibfnamefont{Y.}~\bibnamefont{Ren}},
  \bibinfo{journal}{Phys. Rev. Lett.} \textbf{\bibinfo{volume}{92}},
  \bibinfo{pages}{115001} (\bibinfo{year}{2004}), ISSN
  \bibinfo{issn}{0031-9007}.

\bibitem[{\citenamefont{Fox et~al.}(2010)\citenamefont{Fox, Porkolab, Egedal,
  Katz, , and Le}}]{fox:2010}
\bibinfo{author}{\bibfnamefont{W.}~\bibnamefont{Fox}},
  \bibinfo{author}{\bibfnamefont{M.}~\bibnamefont{Porkolab}},
  \bibinfo{author}{\bibfnamefont{J.}~\bibnamefont{Egedal}},
  \bibinfo{author}{\bibfnamefont{N.}~\bibnamefont{Katz}}, , \bibnamefont{and}
  \bibinfo{author}{\bibfnamefont{A.}~\bibnamefont{Le}}, \bibinfo{journal}{Phys.
  Plasmas} \textbf{\bibinfo{volume}{17}}, \bibinfo{pages}{072303}
  (\bibinfo{year}{2010}).

\bibitem[{\citenamefont{Roytershteyn et~al.}(2013)\citenamefont{Roytershteyn,
  Dorfman, Daughton, Ji, Yamada, and Karimabadi}}]{roytershteyn:2013}
\bibinfo{author}{\bibfnamefont{V.}~\bibnamefont{Roytershteyn}},
  \bibinfo{author}{\bibfnamefont{S.}~\bibnamefont{Dorfman}},
  \bibinfo{author}{\bibfnamefont{W.}~\bibnamefont{Daughton}},
  \bibinfo{author}{\bibfnamefont{H.}~\bibnamefont{Ji}},
  \bibinfo{author}{\bibfnamefont{M.}~\bibnamefont{Yamada}}, \bibnamefont{and}
  \bibinfo{author}{\bibfnamefont{H.}~\bibnamefont{Karimabadi}},
  \bibinfo{journal}{Physics of Plasmas} \textbf{\bibinfo{volume}{20}},
  \bibinfo{eid}{061212} (\bibinfo{year}{2013}),
  \urlprefix\url{http://scitation.aip.org/content/aip/journal/pop/20/6/10.1063%
/1.4811371}.

\bibitem[{\citenamefont{Dorfman et~al.}(2013)\citenamefont{Dorfman, Ji, Yamada,
  Yoo, Lawrence, Myers, and Tharp}}]{dorfman:2013}
\bibinfo{author}{\bibfnamefont{S.}~\bibnamefont{Dorfman}},
  \bibinfo{author}{\bibfnamefont{H.}~\bibnamefont{Ji}},
  \bibinfo{author}{\bibfnamefont{M.}~\bibnamefont{Yamada}},
  \bibinfo{author}{\bibfnamefont{J.}~\bibnamefont{Yoo}},
  \bibinfo{author}{\bibfnamefont{E.}~\bibnamefont{Lawrence}},
  \bibinfo{author}{\bibfnamefont{C.}~\bibnamefont{Myers}}, \bibnamefont{and}
  \bibinfo{author}{\bibfnamefont{T.}~\bibnamefont{Tharp}},
  \bibinfo{journal}{Geophysical Research Letters}
  \textbf{\bibinfo{volume}{40}}, \bibinfo{pages}{233} (\bibinfo{year}{2013}).


\bibitem[{\citenamefont{Zhou et~al.}(2009)\citenamefont{Zhou, Deng, Li, Pang,
  Vaivads, R{\`e}me, Lucek, Fu, Lin, Yuan et~al.}}]{zhou:2009}
\bibinfo{author}{\bibfnamefont{M.}~\bibnamefont{Zhou}},
  \bibinfo{author}{\bibfnamefont{X.}~\bibnamefont{Deng}},
  \bibinfo{author}{\bibfnamefont{S.}~\bibnamefont{Li}},
  \bibinfo{author}{\bibfnamefont{Y.}~\bibnamefont{Pang}},
  \bibinfo{author}{\bibfnamefont{A.}~\bibnamefont{Vaivads}},
  \bibinfo{author}{\bibfnamefont{H.}~\bibnamefont{R{\`e}me}},
  \bibinfo{author}{\bibfnamefont{E.}~\bibnamefont{Lucek}},
  \bibinfo{author}{\bibfnamefont{S.}~\bibnamefont{Fu}},
  \bibinfo{author}{\bibfnamefont{X.}~\bibnamefont{Lin}},
  \bibinfo{author}{\bibfnamefont{Z.}~\bibnamefont{Yuan}}, \bibnamefont{et~al.},
  \bibinfo{journal}{Journal of Geophysical Research: Space Physics}
  \textbf{\bibinfo{volume}{114}} (\bibinfo{year}{2009}).

\bibitem[{\citenamefont{{Schindler} et~al.}(1988)\citenamefont{{Schindler},
  {Hesse}, and {Birn}}}]{schindler:1988}
\bibinfo{author}{\bibfnamefont{K.}~\bibnamefont{{Schindler}}},
  \bibinfo{author}{\bibfnamefont{M.}~\bibnamefont{{Hesse}}}, \bibnamefont{and}
  \bibinfo{author}{\bibfnamefont{J.}~\bibnamefont{{Birn}}},
  \bibinfo{journal}{\jgr} \textbf{\bibinfo{volume}{93}}, \bibinfo{pages}{5547}
  (\bibinfo{year}{1988}).

\bibitem[{\citenamefont{Hesse and Birn}(1993)}]{hesse:1993}
\bibinfo{author}{\bibfnamefont{M.}~\bibnamefont{Hesse}} \bibnamefont{and}
  \bibinfo{author}{\bibfnamefont{J.}~\bibnamefont{Birn}},
  \bibinfo{journal}{Advances in Space Research} \textbf{\bibinfo{volume}{13}},
  \bibinfo{pages}{249} (\bibinfo{year}{1993}).

\bibitem[{\citenamefont{Liu et~al.}(2015)\citenamefont{Liu, Hesse, and
  Kuznetsova}}]{liu:2015}
\bibinfo{author}{\bibfnamefont{Y.-H.} \bibnamefont{Liu}},
  \bibinfo{author}{\bibfnamefont{M.}~\bibnamefont{Hesse}}, \bibnamefont{and}
  \bibinfo{author}{\bibfnamefont{M.}~\bibnamefont{Kuznetsova}},
  \bibinfo{journal}{Journal of Geophysical Research: Space Physics}
  \textbf{\bibinfo{volume}{120}}, \bibinfo{pages}{7331} (\bibinfo{year}{2015}).

\bibitem[{\citenamefont{Sauppe and Daughton}(2018)}]{sauppe:2018}
\bibinfo{author}{\bibfnamefont{J.~P.} \bibnamefont{Sauppe}} \bibnamefont{and}
  \bibinfo{author}{\bibfnamefont{W.}~\bibnamefont{Daughton}},
  \bibinfo{journal}{Physics of Plasmas} \textbf{\bibinfo{volume}{25}},
  \bibinfo{pages}{012901} (\bibinfo{year}{2018}).

\bibitem[{\citenamefont{Torbert et~al.}(2017)\citenamefont{Torbert, Burch,
  Argall, Alm, Farrugia, Forbes, Giles, Rager, Dorelli, Strangeway
  et~al.}}]{torbert:2017}
\bibinfo{author}{\bibfnamefont{R.}~\bibnamefont{Torbert}},
  \bibinfo{author}{\bibfnamefont{J.}~\bibnamefont{Burch}},
  \bibinfo{author}{\bibfnamefont{M.}~\bibnamefont{Argall}},
  \bibinfo{author}{\bibfnamefont{L.}~\bibnamefont{Alm}},
  \bibinfo{author}{\bibfnamefont{C.}~\bibnamefont{Farrugia}},
  \bibinfo{author}{\bibfnamefont{T.}~\bibnamefont{Forbes}},
  \bibinfo{author}{\bibfnamefont{B.}~\bibnamefont{Giles}},
  \bibinfo{author}{\bibfnamefont{A.}~\bibnamefont{Rager}},
  \bibinfo{author}{\bibfnamefont{J.}~\bibnamefont{Dorelli}},
  \bibinfo{author}{\bibfnamefont{R.}~\bibnamefont{Strangeway}},
  \bibnamefont{et~al.}, \bibinfo{journal}{Journal of Geophysical Research:
  Space Physics}  (\bibinfo{year}{2017}).

\bibitem[{\citenamefont{Rager et~al.}(2018)\citenamefont{Rager, Dorelli,
  Gershman, Uritsky, Avanov, Torbert, Burch, Ergun, Egedal, Schiff
  et~al.}}]{rager:2018}
\bibinfo{author}{\bibfnamefont{A.}~\bibnamefont{Rager}},
  \bibinfo{author}{\bibfnamefont{J.}~\bibnamefont{Dorelli}},
  \bibinfo{author}{\bibfnamefont{D.}~\bibnamefont{Gershman}},
  \bibinfo{author}{\bibfnamefont{V.}~\bibnamefont{Uritsky}},
  \bibinfo{author}{\bibfnamefont{L.}~\bibnamefont{Avanov}},
  \bibinfo{author}{\bibfnamefont{R.}~\bibnamefont{Torbert}},
  \bibinfo{author}{\bibfnamefont{J.}~\bibnamefont{Burch}},
  \bibinfo{author}{\bibfnamefont{R.}~\bibnamefont{Ergun}},
  \bibinfo{author}{\bibfnamefont{J.}~\bibnamefont{Egedal}},
  \bibinfo{author}{\bibfnamefont{C.}~\bibnamefont{Schiff}},
  \bibnamefont{et~al.}, \bibinfo{journal}{Geophysical research letters}
  \textbf{\bibinfo{volume}{45}}, \bibinfo{pages}{578} (\bibinfo{year}{2018}).

\end{thebibliography}
\end{document}